\documentclass{SciPost}

\hypersetup{
    colorlinks,
    linkcolor={red!50!black},
    citecolor={blue!50!black},
    urlcolor={blue!80!black}
}

\usepackage[bitstream-charter]{mathdesign}
\DeclareSymbolFont{usualmathcal}{OMS}{cmsy}{m}{n}
\DeclareSymbolFontAlphabet{\mathcal}{usualmathcal}

\fancypagestyle{SPstyle}{
\fancyhf{}
\lhead{\colorbox{scipostblue}{\bf \color{white} ~SciPost Physics Lecture Notes }}
\rhead{{\bf \color{scipostdeepblue} ~Submission }}

\fancyfoot[C]{\textbf{\thepage}}
}

\usepackage[utf8]{inputenc}
\usepackage[T1]{fontenc}
\usepackage{amsmath}
\usepackage{amsfonts}
\usepackage[version=4]{mhchem}
\usepackage{stmaryrd}
\usepackage{graphicx}
\usepackage[export]{adjustbox}
\usepackage{hyperref}
\graphicspath{ {./images/} }

\usepackage{tikz}
\usetikzlibrary{calc}
\usepackage{tikzit}
\usetikzlibrary{decorations.markings}
\usetikzlibrary{decorations.text}
\usetikzlibrary{decorations.pathreplacing}
\usetikzlibrary{decorations.pathmorphing}
\usepackage{xspace}

\tikzstyle{new style 0}=[fill=white, draw=black, shape=rectangle]
\tikzstyle{new edge style 0}=[snake=coil, draw=black]

\tikzstyle{interm}=[-, line width=1 pt, dashed, draw={rgb,255: red,126; green,126; blue,126}]
\tikzstyle{gluon}=[-, draw=black, line width=1 pt, decorate, decoration={{coil,amplitude=1mm,segment length=1.3mm}}]
\tikzstyle{quark}=[-, draw=black, line width=1 pt, postaction={decorate, decoration={{markings,mark=at position 0.5 with {\arrow[#1]{stealth}}}}}]
\tikzstyle{basic}=[-, line width=1 pt]

\newcommand{\YLE}{Yang-Lee edge\xspace}
\newcommand{\TRW}{T_{\rm RW}}

\begin{document}

\pagestyle{SPstyle}

\begin{center}{\Large \textbf{\color{scipostdeepblue}{
Two lectures on Yang-Lee edge singularity \\and analytic structure of QCD equation of state\\
}}}\end{center}

\begin{center}\textbf{
Vladimir V. Skokov
}\end{center}

\author{Vladimir V. Skokov}

\begin{center}
Department of Physics, 
  North Carolina State University,
Raleigh, NC 27695
\\[\baselineskip]
\href{mailto:vskokov@ncsu.edu}{\small vskokov@ncsu.edu}
\end{center}
\section*{\color{scipostdeepblue}{Abstract}}
\textbf{\boldmath{%
These lecture notes, prepared for the 2024 XQCD PhD, provide an introduction to the analytic structure of an equation of state near a second-order phase transition and its most prominent landmark: the Yang-Lee edge singularity. 
In addition to discussing general properties, the notes review recent theoretical progress in locating the QCD critical point by tracking the trajectory of the Yang-Lee edge singularity.
}}

\vspace{\baselineskip}

\noindent\textcolor{white!90!black}{%
\fbox{\parbox{0.975\linewidth}{%
\textcolor{white!40!black}{\begin{tabular}{lr}%
  \begin{minipage}{0.6\textwidth}%
    {\small Copyright attribution to authors. \newline
    This work is a submission to SciPost Physics Lecture Notes. \newline
    License information to appear upon publication. \newline
    Publication information to appear upon publication.}
  \end{minipage} & \begin{minipage}{0.4\textwidth}
    {\small Received Date \newline Accepted Date \newline Published Date}%
  \end{minipage}
\end{tabular}}
}}
}


\vspace{10pt}
\noindent\rule{\textwidth}{1pt}
\tableofcontents
\noindent\rule{\textwidth}{1pt}
\vspace{10pt}

\section{Introduction and motivation}
\label{sec:intro}

In these lecture notes, I discuss the analytic structure of the QCD
phase diagram focusing on potential phase transitions and critical points.
When I refer to the analytic structure, I am simply addressing the existence and positioning of poles, branch points, and branch cuts of pressure (or, more generally, the partition function) in the complex plane of thermodynamic variables (such as the complex baryon chemical potential in the context of QCD).  The groundbreaking work of Lee and Yang highlighted a profound link between the analytical structure of the equation of state and the phase structure \cite{Yang:1952be,Lee:1952ig}. Their primary focus was on finite volume systems. In these notes, I will take a different approach, predominantly considering the thermodynamic limit with only sporadic mentions of finite volume systems.

In my experience, a fraction of the audience loses interest whenever I talk about complex values of thermodynamic variables.  There is a widespread perception that going to the complex plane  
is a purely academic problem with no real-world application to physical systems. 
That is why it is crucial to address the significance of the analytic structure of the equation of state right from the start. 
Firstly, from a theoretical perspective, we aim to learn about QCD in all possible regimes. 
As an extreme example, it is known that some deformations of QCD might allow for an analytic treatment 
of confinement (see e.g. Ref.~\cite{Cherman:2017tey}). Going to complex values of baryon chemical potential alone is not sufficient to carry out this program, 
but it is complementary as it helps to establish if it is possible to remove the deformation without crossing 
phase transitions, critical points.
Secondly, there is a pragmatic reason,  as far as the QCD is concerned. 
The main tool to study the QCD phase diagram non-perturbatively and  from the first
principles is lattice QCD~\footnote{Note that functional methods have become very prominent recently, see, e.g., Refs.~\cite{Fu:2019hdw,Isserstedt:2019pgx}.}. 
At non-zero {\it real} values of the chemical potential,
the lattice approach suffers from the so-called sign problem 
which prohibits efficient importance sampling and renders the approach unpractical. 
This is the main reason why our knowledge of QCD phase diagram at finite chemical potential 
is limited.
Despite this problem, performing lattice QCD calculations at zero or purely imaginary chemical potential can still yield quite a bit of information. 
Let's discuss the former. The idea is to compute the Maclaurin series expansion of pressure 
\begin{align}
  p(\mu, T) \approx 
  p(0, T) + 
  \frac{\partial p(0,T)}{\partial \mu} \mu  +  
  \frac{1}{2}\frac{\partial^2 p(0,T)}{\partial \mu^2} \mu^2  +  
  \frac{1}{3!} \frac{\partial^3 p(0,T)}{\partial \mu^3} \mu^3  +  
  \frac{1}{4!} \frac{\partial^4 p(0,T)}{\partial \mu^4} \mu^4  + \ldots   
\end{align}
Note that all derivatives here are evaluated at zero baryon chemical potential and therefore, are computable on the lattice! 
The above expansion can be simplified by taking into account the quark-anti-quark symmetry of the QCD partition function 
$p(\mu,T) =  p(-\mu,T)$; it leads to zero odd-order derivatives at zero chemical potential 
and to the expansion involving only even powers of $\mu$:
\begin{align}
  p(\mu, T) \approx 
  p(0, T) + 
  \frac{1}{2}\frac{\partial^2 p(0,T)}{\partial \mu^2} \mu^2  +  
  \frac{1}{4!} \frac{\partial^4 p(0,T)}{\partial \mu^4} \mu^4  + \ldots   
\end{align}
This expansion may potentially allow us to study the function $p(\mu, T)$ at any value of $\mu$.  
To illustrate this, let's consider two simplified models mimicking some of the properties of QCD in high- and low-temperature 
limits. At low $T$, the thermodynamics of QCD is well described by the Hadron Resonance Gas model (a collection of non-interacting hadrons). 
Here, we are only interested in baryonic degrees of freedom, as only they depend on the chemical potential. 
Due to the substantial mass of baryons, we also perform degeneracy expansion (see e.g. Appendix A1.4 of Ref~\cite{Kapusta:2023eix}) and keep only the leading order for illustrative purpose~\footnote{Note that 
this approximation is violated at high values of the baryon chemical potential. When all order of the degeneracy expansion are accounted for, the analytic structure is similar to the that of free quarks with $\mu \to \mu_B$.  }. For the pressure (neglecting chemical potential independent terms), 
we obtain
\begin{align*}
  p^{\rm HRG}(\mu_B,T) = C \cosh(\hat \mu_B) ,
\end{align*}
where $\hat \mu_B = \mu_B/T$ and $C$ is a constant whose form is irrelevant to our discussion.  
Series expansion of $\cosh(x) = \sum_{i=0}^\infty \frac{x^{2i}}{(2i)!}$ is known 
to an arbitrary order, the series converges for any value of $x$. 
This can be tested by evaluating the radius of convergence
\begin{align}
  R = \frac{1}{\lim_{n\to \infty} (c_n)^{1/n}} = \frac{1}{\lim_{n\to \infty} \left(\frac{1}{(2n)!} \right)^{1/n}} = \infty \,.
  \label{eq:R}
\end{align}
Practically, this means that if one wants to compute a value of the function 
$p^{\rm HRG}(\mu_B,T)$ at a some $\mu_B = \mu_B^*$  with precision $\delta$, there exists such an 
$n$ for which the truncated Maclaurin series is an appropriate approximation,
that is $|p^{\rm HRG}(\mu_B^*,T) - C \sum_{i=0}^n \frac{\hat \mu_B^{2i}}{(2i)!}| <\delta $.
It is worth mentioning that the function $\cosh(x)$ is analytic in the complex plane of $x$, thus
it does not have singularities at finite (complex) values of  $x$.    

\begin{figure}
  \begin{center}
    \includegraphics[width=0.3\textwidth]{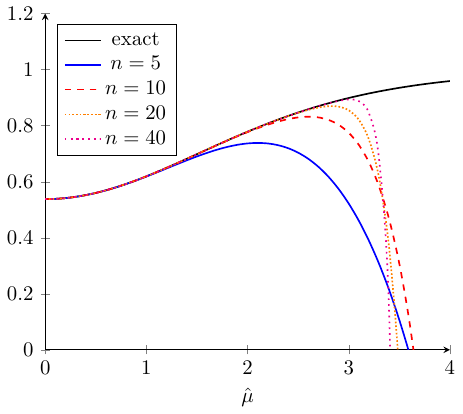}
  \end{center}
  \caption{$n$-th order truncated Maclaurin series approximations vs the function itself for $\hat \omega=1$.}\label{fig:FQexp}
\end{figure}

As the reader suspects, the situation is not that simple in QCD. Our next example of free quarks illustrates this. Consider the corresponding pressure (see Chapter 2 of  Ref~\cite{Kapusta:2023eix})
\begin{align*}
  p^{\rm SB} (\mu,T) = \frac{2 N_c N_f}{3}  \int \frac{d^3 p}{(2\pi)^3} \frac{p^2}{\omega} \frac{1}{e^{\hat \omega - \hat \mu}+1}  + (\mu \to - \mu)
\end{align*}
where $\hat \omega = \sqrt{p^2+m^2}/T$.  
In principle, one can expand this function in $\mu$ around zero and analyze the convergence. 
However, for simplicity's sake, let's consider the integrand only and perform the analysis of the corresponding series expansion. 
We get 
\begin{align}
    \frac{1}{e^{\hat \omega - \hat \mu}+1}  + (\hat \mu \to - \hat \mu)
= \frac{2}{e^{\hat \omega }+1}+2 \sinh ^4\left(\frac{\hat \omega }{2}\right)
   \text{csch}^3(\hat \omega ) \hat \mu^2   +O\left(\hat \mu ^4\right)
   \label{Eq:TaylorSample}
\end{align}
Higher orders can be readily computed, but their analytic form becomes rather cumbersome. 
Instead, I opt for plotting this function and its truncated Maclaurin series approximations 
in Fig.~\ref{fig:FQexp}.
In this figure, we see that beyond a certain value of $\hat \mu$, going to higher expansion orders does not 
improve the description of this function. An analysis of the convergence radius, see Eq.~\eqref{eq:R}, 
gives a finite number $R\approx 3.3$.     
Now, let's analyze this function in the complex plane. It is sufficient to consider only one 
term (Fermi distribution) $f(\hat \mu) =  \frac{1}{e^{\hat \omega - \hat \mu}+1}$. The denominator of this function may vanish at some value of 
$\hat \mu$, resulting in a pole of $f(\mu)$. Simple algebra yields   
\begin{align*}
  \hat \mu^* = \hat \omega \pm i \pi n
\end{align*}
where $n$ is a positive integer. The origin of $n$ is due 
to the periodicity of our function: $f(\hat \mu)$ $= f(\hat \mu+2\pi i)$, which can be proven easily.    
Specifically, the closest singularities to the origin are located at $\hat \mu^* = \hat \omega \pm i \pi$
which, for $\hat\omega = 1$ are the distance $\sqrt{1+\pi^2}\approx 3.3$ away from the origin. 
This is a simple illustration of an important statement:  
{\it the radius of convergence of a series represents
the distance in the complex plane from the expansion point to the nearest singularity of the function expanded}~\footnote{The
generic proof is a byproduct of one of the most important theorems of complex analysis 
``a holomorphic function is analytic and vice versa''.}. 

Before we move on, I want to draw your attention to an interesting fact about the Fermi distribution function when evaluated on the line $\hat \mu = \hat \mu_r + i \pi$: 
\begin{align}
  f(\hat \mu) =  \frac{1}{e^{\hat \omega - \hat \mu_r + i \pi}+1} =  - \frac{1}{1-e^{\hat \omega - \hat \mu_r}} = - f_B(\hat \mu_r), 
\end{align}
where $f_B$ is the Bose distribution and $\hat \mu_r$ is purely real. This equality demonstrates that the Fermi distribution morphs into a
negative Bose distribution which can develop singularity due to Bose-Einstein condensation for $\hat \mu_r = \hat m$.

To summarize, we demonstrated that in the case of   $p^{\rm SB} (\mu, T) $~\footnote{We only analyzed the integrand of this function. The integral will evolve a sum over a continuum of poles. 
This sum leads to a branch point singularity $\hat \mu^* = \hat m \pm i \pi n $  and associated cuts.} what we can learn from the series expansion is limited~\footnote{In Sec.~\ref{Sec:Conformal},  I will return to this and discuss how the knowledge of the analytic structure can help us learn more about the function.}
with singularity in the complex plane defining the horizon of our knowledge.  
Let me stress again: This is the main reason we want to know the analytic structure of QCD.

\section{Analytic structure near a critical point: mean-field approximation}
\label{sec:YLEmf}
The key question is whether phase transitions imprint the analytic structure of the pressure or, in general, the partition function. 
To explore this, we consider Landau's mean-field theory of phase transitions. As far as the long-range order is concerned, the mean-field approximation becomes exact above the upper critical dimension, as I explain in the next section. In application to the physical case of three spatial dimensions, the mean-field theory can only roughly approximate a system near a second-order phase transition. 

\subsection{Landau's mean-field theory}

\begin{figure}[t]
  \begin{center}
    \includegraphics[width=0.3\textwidth]{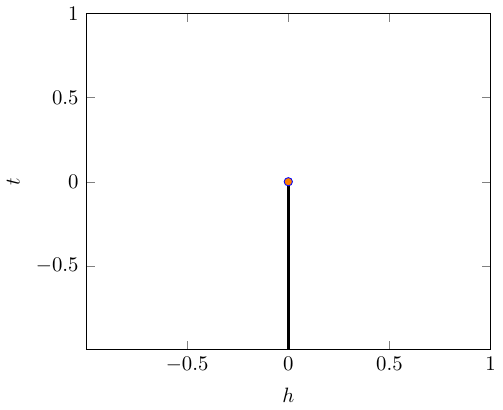}
    \includegraphics[width=0.3\textwidth]{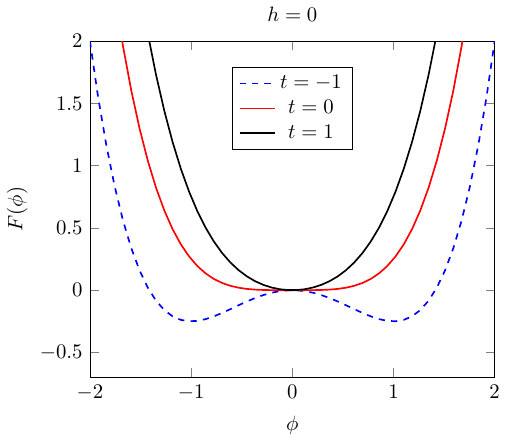}
    \includegraphics[width=0.3\textwidth]{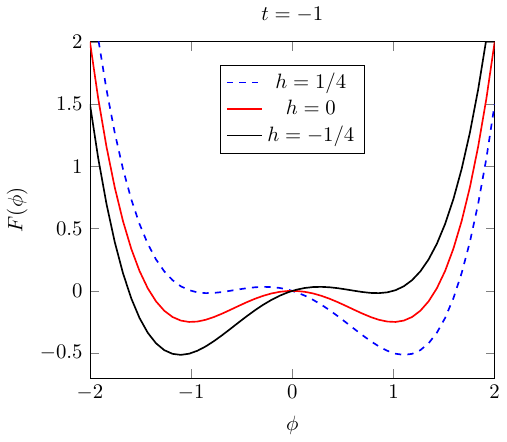}
  \end{center}
\caption{Phase diagram of the mean-field Landau model and two types of phase transitions.
The middle panel illustrates the second-order phase transition at zero external magnetic field $h=0$. 
The right panel shows the first-order phase transition for a fixed negative value of $t=-1$. 
}\label{fig:pd}
\end{figure}

Landau's mean-field model is defined by the following free energy:  
\begin{align}
 F = \frac{t \phi^{2}}{2}+\frac{\lambda \phi^{4}}{4}-h {\phi}\,.
 \label{eq:FL}
\end{align}
To keep the theory stable, here we assume that $\lambda$ is a positive constant.  Therefore an appropriate rescaling of the field $\phi$ and parameters $t$ and $h$ 
is equivalent to setting $\lambda = 1$. 
$t$ is the so-called reduced temperature $t = (T-T_c)/T_c$ and $h$ is the external field.
At zero $h$ the Landau free energy has $Z(2)$ symmetry ($F(\phi) = F(-\phi)$). In contrast to $t$, nonzero values of $h$ break the symmetry 
explicitly.
Analyzing the potential for different values of the parameters allows for easy reconstruction of the model's phase diagram. 
I present an illustration for two different transition scenarios in Fig.~\ref{fig:pd}.
There is a third possible transition scenario,  which in the heavy-ion collision community is known by the name of ``crossover''~\footnote{It is not related to the crossover phenomenon (see Ref.~\cite{Amit:1984ms} for details) of critical statics.}. It denotes a smooth transition at a fixed positive value of $t$ as $h$ changes sign.

The expectation value of the order parameter $\phi$ can be found by minimizing $F$. Before attacking most generic 
case,  let's first consider two special cases: zero $h$ and zero $t$.
\begin{itemize}
  \item At zero $h$, the minimization of the free energy  leads to the following values of the order parameter:
    \begin{equation}
      \phi  = 
      \left\{\begin{array}{lr}
        0, & \text{for } t \ge  0,\\
        \pm(-t)^{1/2}, & \text{for } t<0. 
        \end{array}\right. 
      \label{eq:zeroh}
    \end{equation}
This shows that (in infinite volume) the symmetry is  spontaneously  broken (the system has either a positive or negative value of the order parameter) 
when $t<0$ and the order parameter develops a non-zero expectation value. Generically 
we have $\phi = (-t)^\beta$, where $\beta$ is a universal critical exponent (in Landau theory, $\beta = 1/2$).    
\item At zero $t$, minimization of $F$ leads to 
  \begin{equation}
    \phi = h ^ {1/3}. 
    \label{eq:zerot}
  \end{equation}
 At critical temperature $t=0$,  the dependence of the order parameter on the external field defines  
  the critical exponent $\delta$: $\phi = h ^ {1/\delta}$.  
\end{itemize}

\subsection{Singularities of Landau mean-field theory}

We are ready to analyze the equation of motion in the general case:
\begin{align}
  t \phi + \phi^3 - h = 0\,.
  \label{eq:op}
\end{align}
Here we specifically want to investigate if the order parameter has a singularity in the complex plane 
of  $t$ or $h$. For the sake of the argument, let's fix $t>0$ and consider 
$\phi$ as a function of $h$. The analysis can also be performed for complex values of $t$, but it will not provide any additional insights, as it will be clear below. 

In Eq.~\eqref{eq:zerot}, we saw that at the singularity/critical point the order parameter behaves as $\phi \propto h^{1/\delta}$, where $|\delta|>1$. Generically, if a singularity is not at a zero magnetic field, we have  $\phi \propto (h-h_c)^{1/\delta}$. Let's consider this as an Anzatz. 
Then at $h=h_c$, we get  
\begin{align*}
  \frac{d \phi }{d h } = \infty 
\end{align*}
or equivalently 
\begin{align*}
  \frac{d h }{d \phi } = 0\,. 
\end{align*}
However, there is a more convenient form of this condition. To derive it, consider the 
equation of motion in the implicit form and differentiate it with respect to $h$ to get 
\begin{align}
  \frac{d} {dh} \frac{\partial F}{\partial \phi} = 
  \frac{\partial^2 F}{\partial \phi^2}  \frac{d\phi}{dh} +  
  \frac{\partial^2 F}{\partial \phi \partial h} = 0. 
\end{align}
We have $\frac{\partial^2 F}{\partial \phi \partial h} = -1 $ and thus 
\begin{align}
  \frac{\partial^2 F}{\partial \phi^2}  \frac{d\phi}{dh} = 1   
\end{align}
inferring that the location of the singularity can be found by requiring the vanishing mass of  the order  parameter:
\begin{align}
   \frac{\partial^2 F}{\partial \phi^2} =  t + 3 \phi^2 = 0.  
   \label{eq:zerom}
\end{align}
The easiest way to proceed is to multiply this equation by $\phi/3$ and subtract the result from the equation of motion to get 
\begin{align}
  \phi_c = \frac32 \frac{h_c}{t}.  
  \label{eq:phic}
\end{align}
Substituting this into Eq.~\eqref{eq:zerom} leads to 
\begin{align}
  h_c = \pm i \frac{2}{3 \sqrt{3}} t^{3/2}.  
  \label{eq:hc}
\end{align}
It is worth pausing here to contemplate the consequences of this result. 
First, we found that, indeed, for positive $t$,  there are singularities in the complex $h$ plane; the singularities are located at purely imaginary values of $h$. Second, we found that a simple combination $h / t^{3/2}$ 
or equivalently $z = t / h^{2/3}$ is a pure number ($z_c = \frac{3}{2^{2/3}} e^{\pm i \pi/3}$) independent of $t$ at the singularity. This hints at a certain symmetry in the equation of motion, which we will now identify.

\subsection{Magnetic equation of state}

Let's revise the equation of motion by using the following ansatz for  the solution $\phi = h ^{1/\delta} f_G(t,h)$ with $\delta=3$: 
\begin{align*}
  t h ^{1/3} f_G + h f_G^3  -h &= 0, \\ 
  \frac{t}{h^{2/3}} f_G + f_G^3 - 1 &=0.   
\end{align*}
This demonstrates that $f_G$ is a function of a combination  $z = t/h^{2/3}$ known as a {\it scaling variable}.
The equation defining $f_G$ is called the magnetic equation of state. For a general universality class 
and beyond mean-filed approximation,  one can prove that the scaling variable is given by $z = t / h^{\frac{1}{\beta \delta}}$, 
where $\beta$ and $\delta$ are the critical exponents of the corresponding universality class. Moreover, in contrast to the mean-field results,
the magnetic equations of state are not known analytically (with rare exceptions, e.g., infinite $N$ limit).

\begin{figure}[tb]
  \begin{center}
    \includegraphics[width=0.3\textwidth]{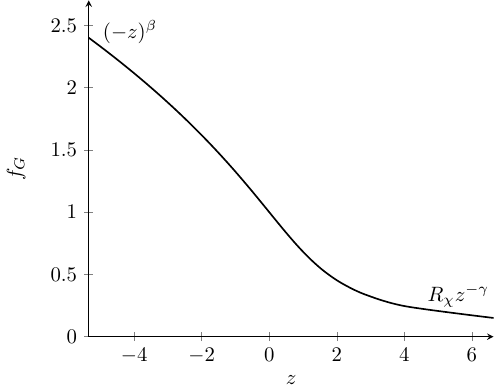}
  \end{center}
\caption{The magnetic equation of state in the mean-field approximation. }\label{fig:meos}
\end{figure}

Figure \ref{fig:meos}
shows the magnetic equation of state (note that you do not have to solve for $f_G$ to plot it; instead, plot the inverse function $z = 1/f_G - f_G^2$).
The conventional way to properly define the magnetic equation of state is to rescale $t$ and $h$~\footnote{This is achieved by introducing the so-called metric factors, see e.g., Ref.~\cite{Engels:2011km} for details} in such a way as to satisfy 
\begin{align}
  f_G(z=0) &= 1, \\
  f_G(z\to - \infty) & = (-z)^\beta. 
\end{align}
Our mean-field result already satisfies these requirements (if we considered $\lambda\ne 1$, it would not be the case). 
The asymptotic behavior of $f_G$ at large positive value of the argument is governed by $R_\chi z^{-\gamma}$, where $R_\chi$ 
is a universal amplitude ratio (in mean-field, $R_\chi=1$) and $\gamma$~\footnote{Note that there are only two independent critical exponents for the critical point. There is an exact (scaling) relation between $\beta, \delta$ and $\gamma$:
  $\gamma = \beta (\delta -1)$. The origin of this and other scaling relations can be explained by the renormalization group; see Ref.~\cite{Zinn-Justin:1996khx, Amit:1984ms} for a detailed introduction.   
} is a universal susceptibility critical exponent (in the mean-field, $\gamma=1$).  

\begin{figure}
  \begin{center}
    \includegraphics[width=0.6\textwidth]{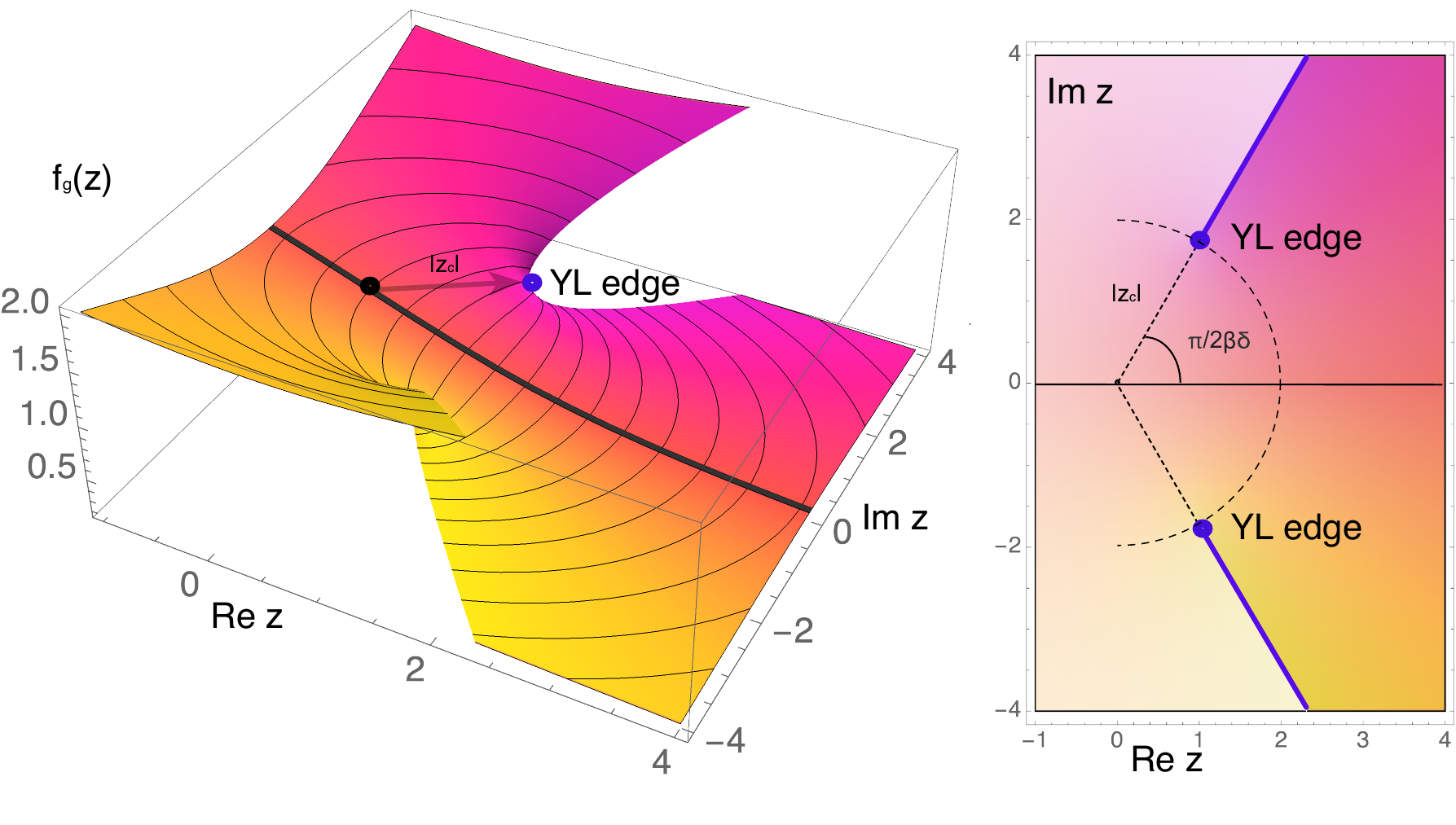}
  \end{center}
  \caption{The real part of $f_G(z)$ as a function of complex $z$. The black solid line in the figure on the left panel is the magnetic equation of state for real values of $z$; see also Fig.~\ref{fig:meos}.  }\label{fig:complexfg}
\end{figure}

As we established, function $f_G$ has two complex conjugate singularities (and associated cuts) in the complex $z$ plane. They are illustrated in Fig.~\ref{fig:complexfg}. 

These are the Yang-Lee edge singularities~\cite{Fisher:1978pf}. 
The first part of the name is due to an intimate connection between these singularities and Lee-Yang zeroes~\footnote{Note that the order of the names is switched. 
Together, Lee and Yang published two papers on the subject, and the papers had different author orderings. } 
introduced in Refs.~\cite{Yang:1952be} and \cite{Lee:1952ig}.
The second part of the name -- ``edge'' refers to the fact that in the finite volume, the cut splits into a countable number of poles of $f_G$ 
with the first pole approximating the location of singularity. Therefore, the singularity defines the edge where the locus of poles originates (or terminates, if you prefer). 

\begin{figure}[b]
  \begin{center}
    \includegraphics[width=0.3\textwidth]{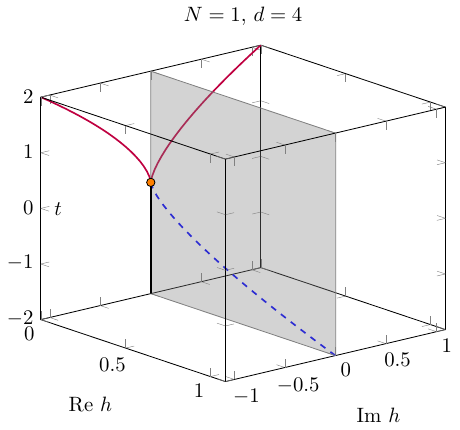}
    \includegraphics[width=0.3\textwidth]{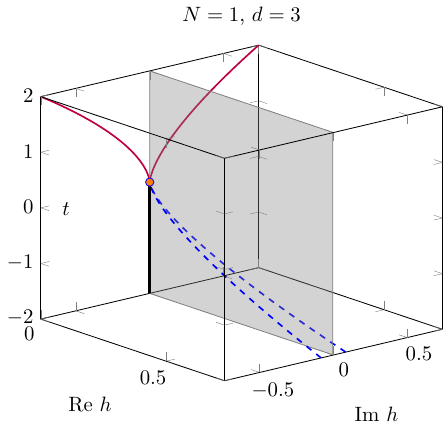}
  \end{center}
\caption{Phase diagram of Ising model in mean-field approximation (or $d=4$) and with fluctuations taking into accounts.}\label{fig:complexPD}
\end{figure}

It is instructive to return to Eq.~\eqref{eq:hc}, and plot a three-dimensional phase diagram by considering real and imaginary values of $h$, see Fig.~\ref{fig:complexPD}.
For now, let's focus on the mean-field result (left panel) and the ``crossover'' region: $t>0$.   
In the three-dimensional diagram, the Yang-Lee edge singularities form lines (each point of the lines is a \YLE). These lines cross at the critical point. In that sense, the critical point has higher co-dimension than the locus of Yang-Lee edge singularities.   Most importantly, Fig.~\ref{fig:complexPD} demonstrates  
that {\it Yang-Lee edge singularity is continuously connected to the critical point}. Thus, tracing the singularity allows one to locate the corresponding critical point.

\subsection{Investigating the Yang-Lee edge singularity}
To what type of singularity does a \YLE  belong? In Figure~\ref{fig:complexfg}
we already saw that  this is a branch point. Let's demonstrate this explicitly. 
We start from the equation for the expectation value of the order parameter~Eq.~\eqref{eq:op}, and we expand it near the Yang-Lee edge: 
\begin{align}
  \phi &= \phi_c  +  \tilde \phi
\end{align}
where $\phi_c$ we found previously in Eq.~\eqref{eq:phic}
and we consider $\tilde \phi/ \phi_c \ll 1$.  
We obtain (after collecting terms)
\begin{align}
  \underbrace{t \phi_c  + \phi_c^3}_{=h_c} + \underbrace{(t + 3 \phi_c^2)}_{=0} \tilde \phi + 3 \phi_c \tilde \phi^2 + \tilde \phi^3 - h =0.
\end{align}
Where the first term gives $h_c$ due to equations of motion, while the second term is zero due to zero mass term at Yang-Lee edge, see Eq.~\eqref{eq:zerom}.
Finally, we can safely neglect $\tilde \phi^3$ term. We thus end up with a simple result: 
\begin{align}
  \tilde \phi &=  \left(\frac{h-h_c}{3\phi_c}\right)^{1/2},  \\
  \phi &= \phi_c + \left(\frac{h-h_c}{3\phi_c}\right)^{1/2}\,.
\end{align}
This shows that, in the mean-field approximation, the  Yang-Lee edge is an algebraic branch point. 
The behavior of the order parameter near the singularity defines the edge critical exponent $\sigma$. 
Thus, as we just showed, in the mean-field, $\sigma_{\rm MF}  = 1/2$. 
The edge singularities can be seen as critical points themselves. As shown in Fig.~\ref{fig:complexPD}, by changing only one parameter, $h$, we can adjust the system to these critical points. In contrast, the conventional critical point requires tuning two parameters, $t$ and $h$ (relevant variables). One would need to adjust four or more parameters to reach a multi-critical point, such as a tricritical point. For this reason, M. Fisher dubbed \YLE singularities {\it protocritical} points. The number of relevant variables also determines the number of independent critical exponents in the leading order scaling relations~\footnote{In these notes, we only consider the minimal set of critical exponents that are required for the narrative. To learn more about a complete set of them and the algebraic relations connecting them, see Ref.~\cite{Amit:1984ms}. }. At the conventional critical point, two relevant perturbations result in two independent critical exponents. On the other hand, at the \YLE, we only have one independent critical exponent, $\sigma$.

Note that from the definition of $\sigma$, one sees its relation to $\delta_{\rm YLE} = 1/\sigma_{\rm YLE}$. It is important to stress that $\delta$ of the underlying universality class does {\it not} have to be (and often not) equal to $\delta_{\rm YLE}$. As the reader might have correctly guessed, the Yang-Lee edge belongs to a different universality class from the conventional critical point of the underlying theory. We will discuss this in the next section. Before we proceed, let's summarize our findings.

\subsection{A brief summary}
\begin{itemize}
  \item Yang-Lee edge singularity is continuously connected to the critical point. Thus, by tracing the trajectory of the singularity as a function of temperature, one can establish the existence and the location of the corresponding critical point.  
  \item Yang-Lee edge is an algebraic branch point singularity. 
  \item When expressed in terms of the scaling variable, the location of the singularity is universal. 
  \item This singularity limits (when it is the closest to the expansion point) the radius of convergence of a series. 
\end{itemize}

There are some questions one might have concerning the list above. 
\begin{itemize}
  \item Would the existence of Yang-Lee edge singularity infer the existence of the corresponding critical point? 
    The answer to this question is a qualifying no. Let me explain by considering a one-dimensional Ising model as an example:
    \begin{align}
      Z = \sum_{\{\sigma\}} e^{- \frac{{H}(\sigma)}{T}} = \sum_{\sigma} e^{\hat{J} \sum_{i=1}^{L-1} \sigma_i \sigma_{i+1} + \hat{h} \sigma_i }\,.
      \label{eq:ZIsing}
    \end{align}
    The free energy of the model is known exactly~\cite{reichl1980modern}: 
    \begin{align}
      f = - \lim_{L\to \infty} \frac{T}{L} \ln Z = -  T \ln \left( e^{\hat J} \cosh(\hat h) + \sqrt{ e^{2\hat J} \sinh^2(\hat h) + e^{-2\hat J}  } \right)
      \label{Eq:fIsing}
    \end{align}
    where  the shorthand notation $\hat x = x/T$ was used.  
    We expect to have an algebraic branch point, and thus it is natural to check when the argument of the square root crosses zero: 
    \begin{align} 
      e^{2\hat J} \sinh^2(\hat h_c) = -  e^{-2\hat J}  
    \end{align}
    or
    \begin{align} 
    h_c = \pm i T \arcsin e^{-2 \hat J}\,.  
    \end{align}
    As expected, we obtained two complex conjugate singularities. They coincide/cross when the imaginary part goes to zero, which is only possible for a fixed value of $J$ when $\hat J$ goes to infinity and thus   $T \to 0$.
    However, a critical point at $T=0$ is a quantum one; the classical Ising model is not appropriate for describing a system in this limit. 
    
    We thus see that the existence of the Yang-Lee edge singularities does not indicate the existence of the {\it finite} temperature critical point. However, the opposite statement is true: every 
    finite temperature critical point has the associated pair of Yang-Lee edges.

    As a by-product of our calculation, expanding the free energy around Yang-Lee edge singularity, we obtain: 
    \begin{align}
      f = T \ln \left(e^{\hat J} \sqrt{1-e^{-4 \hat J}}\right)+ T \frac{(1+i) e^{-\hat J}}{(1-e^{-4 \hat J})^{1/4}} ({\hat h-\hat h_c})^{1/2} + {\cal O}((h-h_c)^{3/2})
    \end{align}
    or simply 
    \begin{align}
      f -f_c \propto (\hat h-\hat h_c)^{1/2} + {\cal O}((h-h_c)^{3/2}) \,.
    \end{align}
    Therefore, the order parameter (the magnetization)  is given by 
    \begin{align}
      \phi -\phi_c \propto  (\hat h-\hat h_c)^{-1/2} + {\cal O}((h-h_c)^{1/2}). 
    \end{align}
    We thus established that in one spatial dimension, the edge critical exponent is negative  $\sigma_{d=1} = -1/2$. 
  \item 
    What is the fate of the Yang-Lee edge singularities in the low temperatures phase $t<0$? 
    
    In this regime, Yang-Lee edge singularities are known as spinodals, see also Refs.~\cite{Stephanov:2006dn,An:2016lni}. At a given $t<0$, the spinodal point is defined by the limiting value of $h$ at which the system loses 
    its local stability against phase separation concerning small fluctuations. Figure~\ref{fig:spinodal} 
    illustrates the system close and at the spinodal point. 
    \begin{figure}
      \begin{center}
        \includegraphics[width=0.3\textwidth]{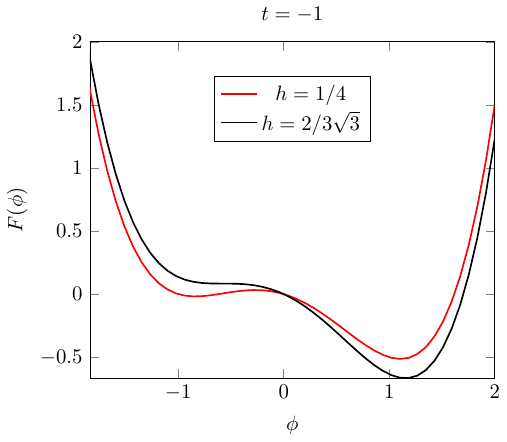}
      \end{center}
      \caption{The Landau potential close to and at the spinodal point.}\label{fig:spinodal}
    \end{figure}
    Note that spinodal lines lie on the real $h$ axis only in the mean-field approximation. Recall that  we have 
    \begin{align}
       h_c = \pm i |z_c|^{-\beta\delta} t^{\beta \delta}\,.
    \end{align}
    For negative $t$ and non-integer or non-half-integer $\beta\delta$, $ i t^{\beta\delta}  = i |t|^{\beta\delta} e^{\pm i \pi \beta\delta}$ is a complex number. However, for 
    half-integer $\beta \delta$ (as we have in mean-field), this combination becomes purely real; see Fig.~\ref{fig:complexPD} for illustration.
    Thus, as far as the {\it real} values of the thermodynamic parameters are concerned, a spinodal is a mean-field object and does not exist in ``real'' systems in thermodynamic equilibrium. 

    \end{itemize}

    \section{Beyond mean-field approximation}
    \subsection{Upper critical dimension}
    The previous section discussed most of the system in the mean-field approximation framework (except for the one-dimensional Ising model). However, going beyond it and accounting for fluctuations is not a simple task in general but is often essential to describe a system near a critical point.
    The role of fluctuations depends sensitively on the number of spatial dimensions. I will motivate below that
    there exists such a $d_c$  - the dimension where fluctuations become important -  above which $d\ge d_c$ the mean-field approximation is sufficient to describe long-range order. 
    $d_c$ is usually referred to as the upper critical dimension. Our first goal is to establish the upper critical dimension in the vicinity of the critical point and Yang-Lee edge singularity.

    Once we account for fluctuations, most of the universal properties are not computable analytically. 
    Moreover, in contrast to mean-field approximation when the number of spatial dimensions, $d$ and the number of the field components, \( N \) 
    never explicitly entered into any of our calculations, universal properties (e.g., critical exponents, the position of the singularity $z_c$) depends on (and often 
    {\it only} on) $d$ and $N$.~\footnote{The edge critical exponent $\sigma$ is independent of $N$ of the underlying universality class. This is due to explicit symmetry breaking at non-zero values of $h$, which singles out one critical mode. $\sigma$ does depend on the number of dimensions as we already showed that $\sigma_{d=1}=-1/2$ and $\sigma_{d\ge 6} = 1/2$. } 

    Let's consider our theory near the Yang-Lee edge to motivate further discussion. We start with the Ginzburg-Landau Lagrangian 
    \begin{align}
      L = (\nabla \phi)^2 + \frac{1}{2} t \phi^2 + \frac{\lambda}{4} \phi^4 - h \phi\,.  
    \end{align}
    This Lagrangian describes the vicinity of the critical point. We can now shift the field variable $\phi = \phi_c + \varphi$. 
    Performing the expansion, we get 
    \begin{align}
      L = (\nabla \varphi)^2 + \frac{1}{2} (t + 3 \lambda \phi_c^2 ) \varphi^2  - \lambda \phi_c \varphi^3   - (h + t\phi_c + \lambda \phi_c^3 ) \varphi\,.  
    \end{align}
    Here, I neglected the $\varphi^4$ as it is irrelevant given the presence of the cubic interaction term. Now we can fix $\phi_c$ by requiring the absence of the mass term (i.e., tune the system to the vicinity of the \YLE; note that for $t>0$, zero mass leads to purely imaginary $\phi_c$). 
    We get
    \begin{align}
      L \approx (\nabla \varphi)^2  - \lambda \phi_c \varphi^3   - H \varphi  
    \end{align}
  with $H = h + t \phi_c + \lambda \phi_c^3$.
  I want to highlight a few interesting observations about the obtained Lagrangian. First, the coupling constant $-\lambda \phi_c$ is purely imaginary due to the presence of the factors $\phi_c$. Second, and most important, we obtained a $\varphi^3$ theory (not $\phi^4$). To understand why it matters, we must take a short detour into the QFT renormalization theory. 
  
In QFT, theories are classified into non-renorma\-lizable, renormalizable, and super-renorma\-lizable. The theory is renormalizable if the coupling constant is dimensionless~\cite{Peskin:1995ev}. In QFT, the renormalization technique is used to treat UV behavior. In contrast, the theory of critical phenomena 
   is concerned with infrared behavior. Below, we demonstrate how these regimes are connected.

  Let's determine the canonical dimension of the coupling constant in a theory with interaction $\lambda_r \phi^r$ in $d$ dimension.  
  The field $\phi$ has dimension of $\Lambda^{\frac{d-2}{2}}$, where $\Lambda$ denote units of inverse length (or momentum). This can be easily determined by considering the kinetic term and remembering that the action is dimensionless (in natural units $c=\bar h = 1$).
  For the coupling constant we thus have $[\lambda_r] \Lambda ^ {\frac{d-2}{2} r } \Lambda^{-d} = 1$ or $[\lambda_r] = \Lambda^{d+r-\frac{1}{2} r d }$  .
  Therefore, the coupling constant of $\lambda \phi^4$ theory is dimensionless in $d=4$, while the one of  $\lambda \phi^3$ theory in $d=6$. The theories are also renormalizable in the same number of dimensions.

  For $\phi^4$ interaction, the theory is super-renormalizable for $d<4$ from the QFT perspective, while from the standpoint of critical phenomena, this is where the theory requires the most complex calculations. 
  Vice versa, for $d>4$, the theory is non-renormalizable in QFT, but in statistical physics, its description is the simplest: Gaussian approximation/mean-field.

  To motivate why this is the case, let's consider a simple graph in $\phi^3$ theory~\footnote{This part is based on  Ref.~\cite{Amit:1984ms}.}:
  \begin{align}
    I(k) = \begin{tikzpicture}[baseline={([yshift=-.8ex] current bounding box.center)}]
	\begin{pgfonlayer}{nodelayer}
		\node [style=none] (0) at (-13, 9) {};
		\node [style=none] (1) at (-11, 9) {};
		\node [style=none] (2) at (-6, 9) {};
		\node [style=none] (3) at (-4, 9) {};
		\node [style=none] (4) at (-12.5, 8.25) {$k$};
		\node [style=none] (5) at (-8.5, 10.75) {$q$};
		\node [style=none] (6) at (-8.5, 7) {$q-k$};
	\end{pgfonlayer}
	\begin{pgfonlayer}{edgelayer}
		\draw [style=quark] (0.center) to (1.center);
		\draw [style=quark, bend left=45, looseness=1.25] (1.center) to (2.center);
		\draw [style=quark] (2.center) to (3.center);
		\draw [style=quark, bend right=45, looseness=1.25] (1.center) to (2.center);
	\end{pgfonlayer}
\end{tikzpicture}
 = \int d^d q \frac{1}{(q^2 + t) ((k-q)^2+t)} \,.
  \end{align}
Field theoretically, $t$ is the mass squared. We want to analyze the behavior of this integral in the infrared by rescaling the external momentum and the mass squared $k\to \alpha k$ and $t\to \alpha^2 t$ with $\alpha\to 0$.
For $d>4$, $I(k)$ is UV divergent, and although we are not interested in its UV structure, it presents a technical difficulty for analyzing the IR. To isolate IR, we will perform the following trick.   
First, we insert 
\begin{align}
  \frac{1}{d} \sum_{i=1}^d \frac{\partial q_i}{\partial q_i} = 1   
\end{align}
into the integrand. Second, we perform the integration by parts. This will give two distinct contributions 
\begin{align}
  I(k) = \frac{1}{d} \int_S \frac{\vec{q}\cdot d\vec{s}}  { (q^2 + t) ((k-q)^2+t) }
  + \frac{1}{d}  \int d^d q \frac{2}{(q^2 + t) ((k-q)^2+t)}  \left[ \frac{q^2}{q^2 + t } - \frac{\vec{q}\cdot (\vec{k}-\vec{q})} { (k-q)^2+t} \right]\,.
\end{align}
The first term is a surface contribution evaluated in the UV, $|q|\sim \Lambda$  (UV cutoff). This contribution is analytic for vanishing $k$ and $t$. We can safely drop it.   
The remaining contribution can be rewritten as 
\begin{align}
  I(k) = \frac{2}{d} \left[ 2 I(k) -  \int  \frac{1}{(q^2 + t) ((k-q)^2+t)}  \left( \frac{t}{q^2 + t } + \frac{\vec{k}\cdot (\vec{k}-\vec{q})+t} { (k-q)^2+t} \right)  \right] 
\end{align}
or 
\begin{align}
  I(k) = \frac{2}{4-d} \left[  \int  \frac{1}{(q^2 + t) ((k-q)^2+t)}  \left( \frac{t}{q^2 + t } + \frac{\vec{k}\cdot (\vec{k}-\vec{q})+t} { (k-q)^2+t} \right)  \right] .
\end{align}
This integral is now convergent for $d<5$. This procedure can be continued until one ends up with UV convergent integrals in which one can set $\Lambda \to \infty$, thus removing an extra scale. The resulting expression will be a homogeneous function of $\alpha$ (since the only present scales are $k$ and $t$). The degree of the homogeneity of the graph is defined by the graph's dimension, $\mathfrak{d}$. Its calculation is identical to the one in QFT; thus 
I will only briefly outline it.    

Again, we will consider $\lambda_r \phi^r$ interaction in $d$-dimensions and focus on computing the dimension of $n$-th order vertex with $E$ external legs. 
The graph behaves as $\alpha^\mathfrak{d}$ with $\mathfrak{d} = Ld - 2 I$, where $L$ is the number of loops (loop integrations) and $I$ is the number of the internal lines (dimension of each propagator is $-2$). 
We can express the number of loops $L$ as the number of internal lines minus the number of momentum conservation conditions. Each interaction brings a momentum-conserving conserving delta-function, 
but one of them should not be included as it is responsible for the overall momentum conservation, thus $L = I - (n-1)$. The number of the internal lines is given by the number of 
all lines from the interaction vertices ($n r$) minus the number of the connected to the external legs. However, it is important to remember that  two lines form a propagator, thus 
$I  = \frac{nr - E}{2}$. Substituting these facts into $\mathfrak{d}$, we get 
\begin{align}
  \mathfrak{d} = - n \left(r+d-\frac{1}{2} r d \right) + \left(d + E - \frac{1}{2} Ed \right).
  \label{eq:grdim}
\end{align}

The question is for what value of $  \left(r+d-\frac{1}{2} r d \right)$ one expects lower scaling $\mathfrak{d}$ with rising complexity/order of the graph $n$? The answer is for $ \left(r+d-\frac{1}{2} r d \right) <0$, so that the first term in
Eq.~\eqref{eq:grdim} is positive. In this case, larger orders $n$ become less significant (due to the suppression $\alpha^\mathfrak{d}$ as $\alpha\to 0$) in the IR, and there is no need to consider complex/large $n$ graphs! 
We thus conclude that the theory is simplest for 
\begin{align}
 r+d-\frac{1}{2} r d  \le 0  \quad \leadsto \quad 
  d \ge  d_c = \frac{2 r}{r -2}\,. 
  \label{eq:dc}
\end{align}

For an illustration, let's consider a two-point function of  $\phi^4$ theory. In $d=4$, one gets $\mathfrak{d} =  2$ (nor relevant in IR) at any order $n$. Thus, the free theory will give the leading infrared behavior.   
In $d=3$,   $\mathfrak{d} = - n  + 2$; thus, the higher order graphs play an increasingly important role in the IR. The theory is non-classical and would require the calculation of arbitrary complex graphs.
Note that the condition~\eqref{eq:dc} coincides with the one from the dimensionality of the coupling constant!   

Having learned that, let's return to the discussion of critical points and Yang-Lee edges. We found that near the critical point ($\phi^4$) the theory has an upper critical dimension of four, while near Yang-Lee edge singularity ($\phi^3$), it is 6! This has severe implications, which we will discuss next!  

\subsection{Methods for including  fluctuations}
Our goal is to locate the Yang-Lee edge singularity in a full theory, including the fluctuations in $d=3$. What methods are usually applied to study critical phenomena?   
\begin{itemize}
  \item The first method that comes to mind is the $\varepsilon$ expansion. The idea of the method is to consider the system close to but not exactly at the critical dimension. 
    The critical exponents and other universal quantities are independent of $d$  and assume their mean-field values above the upper critical dimension $d_c$.
    Below $d_c$, universal quantities have a nontrivial dimensionality dependence. 
    Specifically, the critical exponents are smooth, analytical functions of the dimensionality $d$. 
    This observation, made by Wilson and Fisher, led to the formulation of a systematic and controlled 
    RG approach based on an expansion in the small parameter $\varepsilon = d - d_c$. For most of the quantities, 
    $\varepsilon$ expansion is the expansion in the number of loops. For example, for the critical exponent $\beta$ (in Ising universality class, $N=1$) we have~\footnote{This can be obtained from $\gamma$ and $\eta$ derived in Ref.~\cite{BREZIN1973227}  and the scaling relation $\beta = \frac{\gamma}{2}\left( \frac{d}{2-\eta} -1\right)$.} 
    \begin{align*}
      \beta = \frac{1}{2} + \underbrace{\frac{1}{6} \varepsilon}_{\text{one loop}} + \underbrace{\frac{1}{162} \varepsilon^2}_{\text{two loop}} + \underbrace{\frac{1}{2} \left( \frac{163}{8748} - \frac{2}{27} \zeta(3) \right) \varepsilon^3}_{\text{three loop}} + {\cal O}(\varepsilon^4), \quad \epsilon = 4 - d 
    \end{align*}
    where the leading order is the result of the mean-field approximation. By performing the series analysis, one can extrapolate the value of $\beta$ to $d=3$ ($\varepsilon=1$).  
    
    The $\varepsilon$-expansion was successful in determining many universal properties. This is why applying this method to computing the universal location of the Yang-Lee edge $z_c$ is reasonable. 
    Here, however, a careful reader might anticipate a problem. As we concluded in the previous subsection, the theory has an upper critical dimension of 6 near Yang-Lee edge, higher than the 
    one of the underlying universality class (4). Thus, it is natural to expect that $\varepsilon$-expansion would fail, as it is designed to work just below the upper critical dimension. 
    This said, nothing prevents one from plowing ahead and compute the location (see Ref.~\cite{Johnson:2022cqv} for details):
    \begin{align}
      |z_c| \approx |z_c^{\rm MF}| \left[1 + \frac{ 27 \ln \left(\frac{3}{2}\right) - (N-1) \ln 2}{9 (N+8)}\varepsilon + \underbrace{\varepsilon^2 \log \varepsilon \times ( \cdots )}_{\text{all loops contribute}} \right]  .
      \label{eq:zcfromeps}
    \end{align}
    Here we see that the first correction to mean-field is of the order of $\varepsilon$ as expected; however, the next correction contains the logarithm of $\varepsilon$, and most importantly, its coefficient cannot be extracted perturbatively as graphs of all orders contribute to it. Thus, the $\varepsilon$-expansion breaks down as it does not provide a controlled way to extract the location of the Yang-Lee edge (see Ref.~\cite{An:2017brc} for a detailed discussion).  
   
  \item The other widely utilized method is Monte-Carlo simulation on a discretized lattice. Monte-Carlo simulations were very successful in providing first principle results 
    on the critical statics 
    in $d=3$. However, the calculations at complex values of parameters are impractical due to the sign problem. For illustration, let's consider the Ising model in one dimension, Eq.~\eqref{eq:ZIsing}.
    Monte-Carlo simulations are based on the importance sampling of the probability distribution, for the Ising model, we have 
    \begin{align}
      P(\sigma) \propto e^{\hat{J} \sum_{i=1}^{L-1} \sigma_i \sigma_{i+1} + \hat{h} \sum_i \sigma_i}.  
    \end{align}
    Consider purely imaginary values of $\hat h = i \theta$. The probability becomes a complex number. This can be partially fixed by considering combining the probability for $\sigma$ and $-\sigma$ configurations simultaneously: 
    \begin{align}
    P(\sigma) \propto e^{\hat{J} \sum_{i=1}^{L-1} \sigma_i \sigma_{i+1}} \cos\left( {\theta \sum_i \sigma_i} \right).  
    \end{align}
    We have an improvement; the probability is no longer a complex number. However, it is not positive semi-definite and thus prevents one from relying on the importance sampling, rendering the approach impractical.  This is the essence of the sign problem (see Ref.~\cite{deForcrand:2009zkb} for an introduction
  in the QCD context).
\end{itemize}

Given the failure of these two methods, we have a limited number of options. One of them is the functional renormalization group approach. 

\subsection{Functional renormalization group approach }
In the previous subsection, for the purpose of locating the \YLE,  we identified two problems with the most common methods applied in studying critical phenomena. 
The Functional Renormalization Group (FRG) approach allows one to circumvent both issues, 
as it does not rely on $\varepsilon$-expansion, nor does it suffer from a sign problem.

We briefly overview the FRG approach; for a thorough review, see Refs.~\cite{Wetterich:1992yh, Berges:2000ew, Delamotte:2003dw, Braun:2011pp, Kopietz:2010zz, Dupuis:2020fhh}.  
The FRG is a specific field-theoretical implementation of Wilson's idea of integrating over momentum shells, which is achieved by including fluctuations ordered by momentum scales. This is done by modifying the path integral measure by adding a mass-like term $\Delta S_k[\varphi]$ suppressing contributions of IR momentum modes with $p\lesssim k$. Under appropriate conditions on $\Delta S_k[\varphi]$, variation of the scale parameter $k$ will lead to an equation connecting the UV effective action at the scale $\Lambda$,  $\Gamma_{k=\Lambda}[\phi] \approx S[\phi]$ to the full IR effective action at $k=0$, $\Gamma_{k=0}[\phi]=\Gamma[\phi]$. The expectation value (or the order parameter) $\phi$ is given by  $\phi(x)=\langle \varphi(x) \rangle$.

In the presence of $\Delta \, S_k[\phi]$, the partition function reads
\begin{eqnarray}
\mathcal{Z}_k[J]&=&\int\mathcal{D}\varphi \, e^{-\Delta S[\varphi]} \,\, e^{-S[\varphi]+J\varphi}
\end{eqnarray}
and thus becomes scale-dependent. Usually, the following choice is considered 
\begin{equation}
    \Delta S_k[\varphi] = \frac{1}{2}\int d^d x \int d^d y \,\, \sum_{i} {\varphi_i(x) R_k(x,y)\varphi_i(y)}. 
\end{equation}
In order to match the symmetry of the system, the mass-like regulator function $R_k(x,y)$ is chosen to be invariant under rotations (including the internal space rotations) and translations, i.e., $R_k(x,y)=R_k(x-y)$. Furthermore, to suppress modes with $p\ll  k$ while leaving modes with $p \gg k$ intact, the following must hold for the Fourier transform of the regulator:
\begin{eqnarray}
R_k(p) &\propto& \, k^2 \hspace{.5cm} \mbox{for} \hspace{.5cm}  p\ll k, \\
R_k(p) &\to& \, 0 \hspace{.5cm}  \mbox{for} \hspace{.5cm}  p\gg k.
\end{eqnarray}
$R_k(p)$ adds a mass of order $k^2$ to the low-energy modes, thereby suppressing their contributions to the path integral.

The  effective action, 
 $\Gamma_k[\phi]$ is obtained via a modified Legendre transform
\begin{eqnarray}
\label{Eq:gibbs}
\Gamma_k[\phi]&=-\mathrm{ln} \, \mathcal{Z}_k[J]+J \phi-\Delta S_k[\phi].
\end{eqnarray}
The functional $\Gamma_k[\phi]$ satisfies, so-called,  Wetterich equation~\cite{Wetterich:1992yh,Morris:1993qb,Ellwanger:1993mw},
\begin{equation}
\label{Eq:flow}
    \partial_k\Gamma_k[\phi]=\frac{1}{2} \mbox{Tr} \Bigg\{ \partial_kR_{k}\bigg(\frac{\delta^2\Gamma_k[\phi]}{\delta \phi_i \delta \phi_j}+R_{k}\bigg)^{-1} \Bigg\}\,,
\end{equation}
also known as the flow equation. It prescribes the behavior of $\Gamma_k$ between the classical tree-level action at an initial scale $k=\Lambda$ in the ultraviolet, $\Gamma_{k=\Lambda}=S$ -- an initial condition, and the desired full quantum action at $k=0$, $\Gamma_{k=0}=\Gamma$. The FRG equation provides a versatile realization of the Wilsonian RG and is well-suited to study critical physics. Both the scaling function and the critical exponents have been computed in great detail for O$(N)$ theories for real external fields with the FRG; see, e.g., Refs.~\cite{Berges:1995mw, Bohr:2000gp, Litim:2001dt, Bervillier:2007rc, Braun:2007td, Braun:2008sg, Benitez:2009xg, Stokic:2009uv, Litim:2010tt, Benitez:2011xx, Defenu:2014bea, Codello:2014yfa, Eichhorn:2016hdi, Litim:2016hlb, Juttner:2017cpr, Roscher:2018ucp, Yabunaka:2018mju, DePolsi:2020pjk}.

While the flow equation is exact, it defines an infinite tower of coupled partial differential equations for the effective action and its functional derivatives. With few exceptions (like the $O(N)$ model in the large-$N$ limit \cite{Tetradis:1995br}), truncation is therefore necessary. Often, there is  no obvious small parameter that can be used to define a systematic truncation scheme. Fortunately, this is not the case for critical physics, where the diverging correlation length facilitates a systematic expansion about vanishing momentum. Such a derivative expansion \cite{Morris:1994ie} has been shown to have a finite radius of convergence. A method to estimate  
the systematic error has also been developed  \cite{Balog:2019rrg, DePolsi:2020pjk}. Currently, only the next-to-leading order of this expansion, i.e.\ first order in momentum-squared, was applied to locating the \YLE, as doing so in the FRG approach is numerically much more challenging than computing the critical exponents at a conventional critical point~\footnote{The critical exponents were computed to the next-to-next-to-leading order of the derivative expansion in Ref.~\cite{DePolsi:2020pjk}.} or the \YLE. 
A brief explanation why: the practical implementation of the FRG boils down to a set of integro-differential equations. 
To determine the critical exponents it is sufficient to find a fixed point of the set of these equations; thus the problem reduces to a set of algebraic equations involving integrals. 
It can be tackled with the Newton-Raphson method and a fast integrator, e.g.\ a fixed order Gauss–Kronrod quadrature. In contrast to this, locating the  \YLE requires solving
the actual set of the integro-differential equations. In order to make this problem manageable on the current hardware, in Refs~.\cite{Rennecke:2022ohx, Johnson:2022cqv}, we used a specific form of the  FRG cut-off function,
that allows for analytical evaluation of the integrals and simplifies the problem to the set of differential equations. The disadvantage of this choice of the regulator is that it 
limits the order of the derivative expansion we can work at (not higher than next-to-leading order). This, in combination with the computational constraints,  has not let us match the precision of our calculation
to that of the critical exponents in Refs.~\cite{Balog:2019rrg, DePolsi:2020pjk}.     

\subsection{Yang-Lee edge with FRG: collection of results}
Here I briefly summarize the results of Refs.~\cite{Rennecke:2022ohx,Johnson:2022cqv}.
At the next-to-leading order of the derivative expansion, the truncation of the effective action reads 
\begin{align}
\notag
\Gamma_k[\phi] = \int d^d x \left( U_k(\phi)  + \frac {1 } { 2 }  Z_k(\phi)  (\partial_i \phi)^2 \right) \,.
\end{align}
The flow equations can be reformulated as a set of equations for the effective potential $U$
and the wave function renormalizations $Z$. 
For the average potential, one gets 
\begin{align*}
    \partial_t U_k(\rho) &= \frac{1}{2} \int \bar d^dq\,   \partial_t R_k\left(q^2\right) \Big[G_k^\|+(N-1) G_k^\perp\Big]\,, \quad \rho = \frac{\phi^2}{2}
\end{align*}
with 
\begin{align*}
&G_k^{\perp} =  \frac{1}{Z_k^{\perp}(\rho)q^2 + U_k'(\rho) + R_k(q^2)}, \quad G_k^{\|} =  \frac{1}{Z_k^{\|}(\rho)q^2 + U_k'(\rho)+2\rho U_k''(\rho) + R_k(q^2)}.
\end{align*}
Here we introduced the longitudinal and the transverse (Goldstone) modes.  

\begin{figure}
    \centering
       \includegraphics[width=0.49\linewidth]{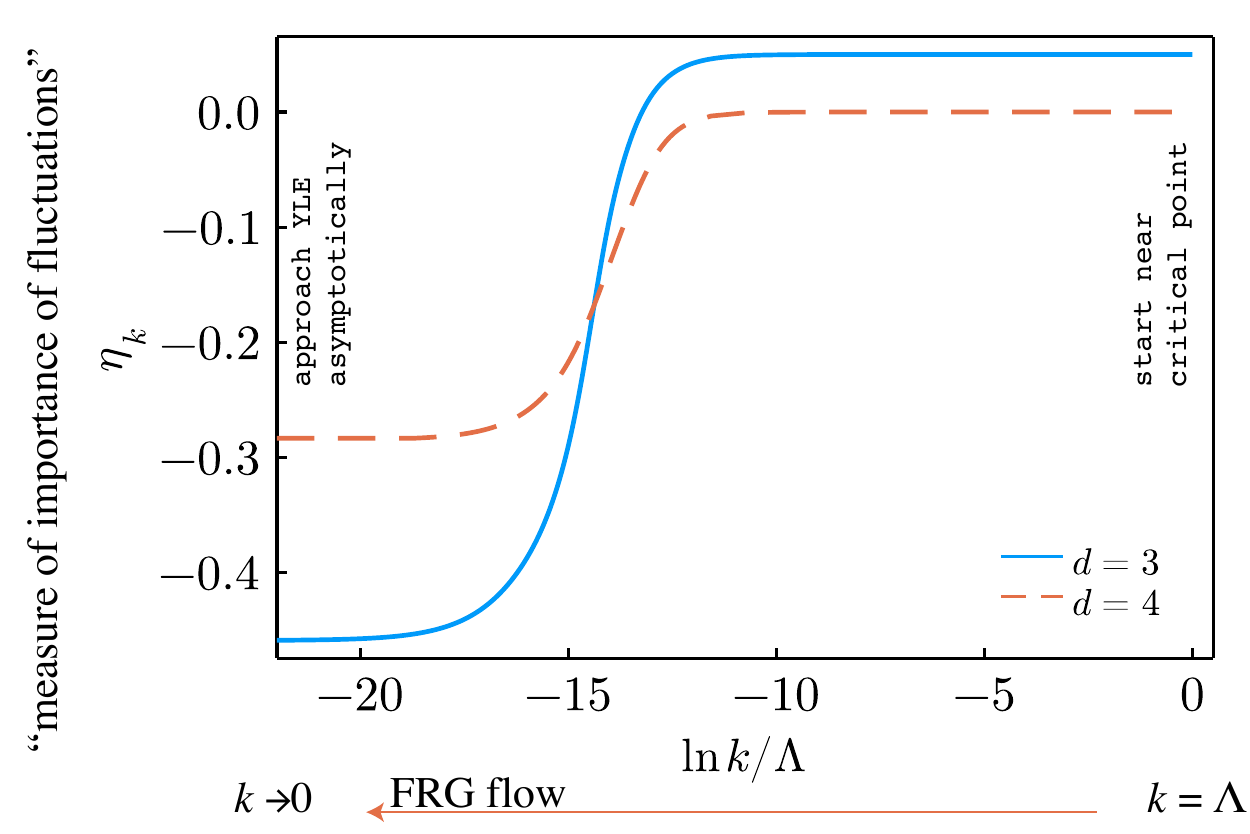}
    \caption{ The effective anomalous dimension as a function of the cut-off momentum $k$ in FRG evolution in $d=3$ and $d=4$. 
    The initial conditions for the FRG flow are formulated in the vicinity of the Wilson-Fisher critical point; the flow equation is constructed in such a way as to track the position of the \YLE singularity  $m_R = 0+$ . In the IR ($k \to 0$), the anomalous dimension reaches its physical value at the \YLE singularity. 
      The magnitude of $\eta$ can be considered as a measure of the importance of fluctuations. In this context, it is interesting to consider $d=4$. Near Wilson-Fisher point $\eta =0$, that is fluctuations can be neglected and the mean-filed approximation becomes exact. As the system is driven towards the \YLE fixed point, the magnitude of the anomalous dimension increases drastically; this illustrates the fact that the upper critical dimension of the \YLE fixed point (6) is larger than the one of the Wilson-Fisher fixed point (4).  
    }
    \label{fig:etaevo}
\end{figure}

\begin{figure}[tb]
    \centering
       \includegraphics[width=0.39\linewidth]{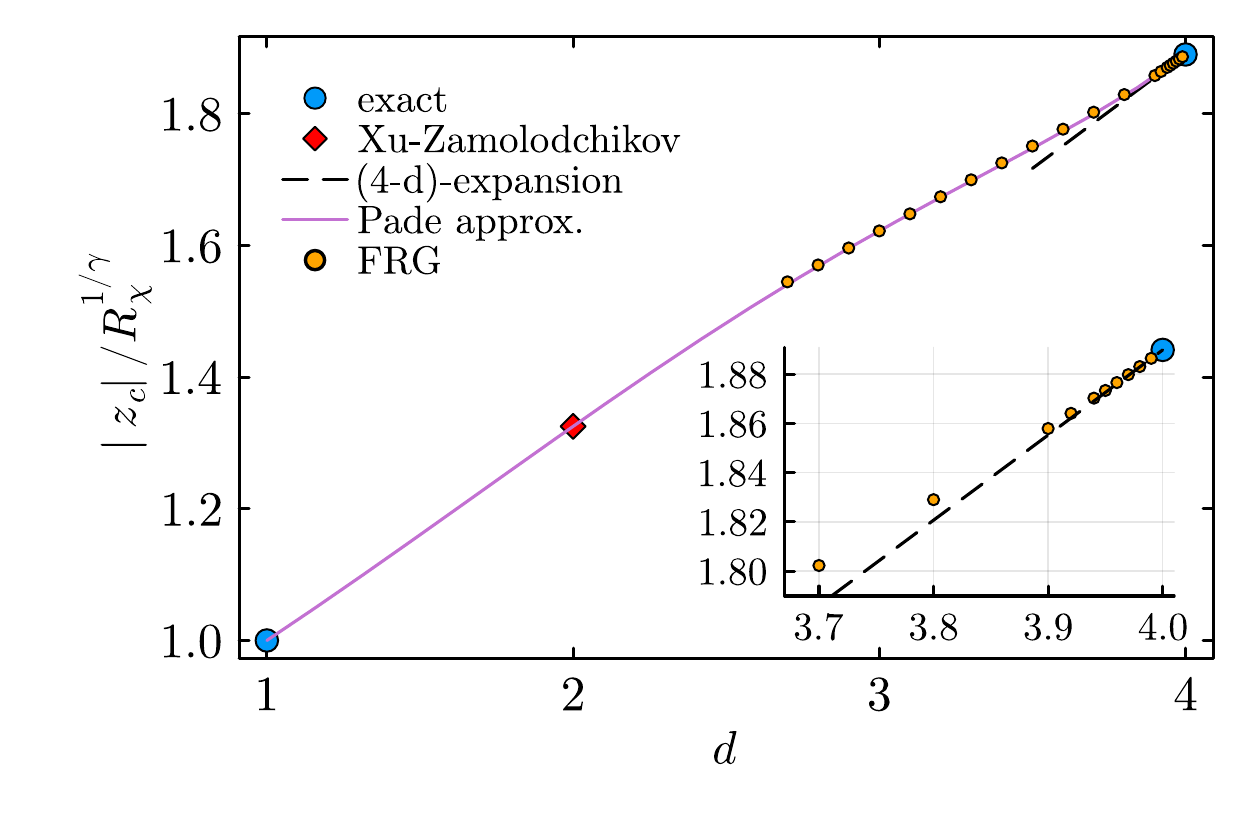}
       \includegraphics[width=0.39\linewidth]{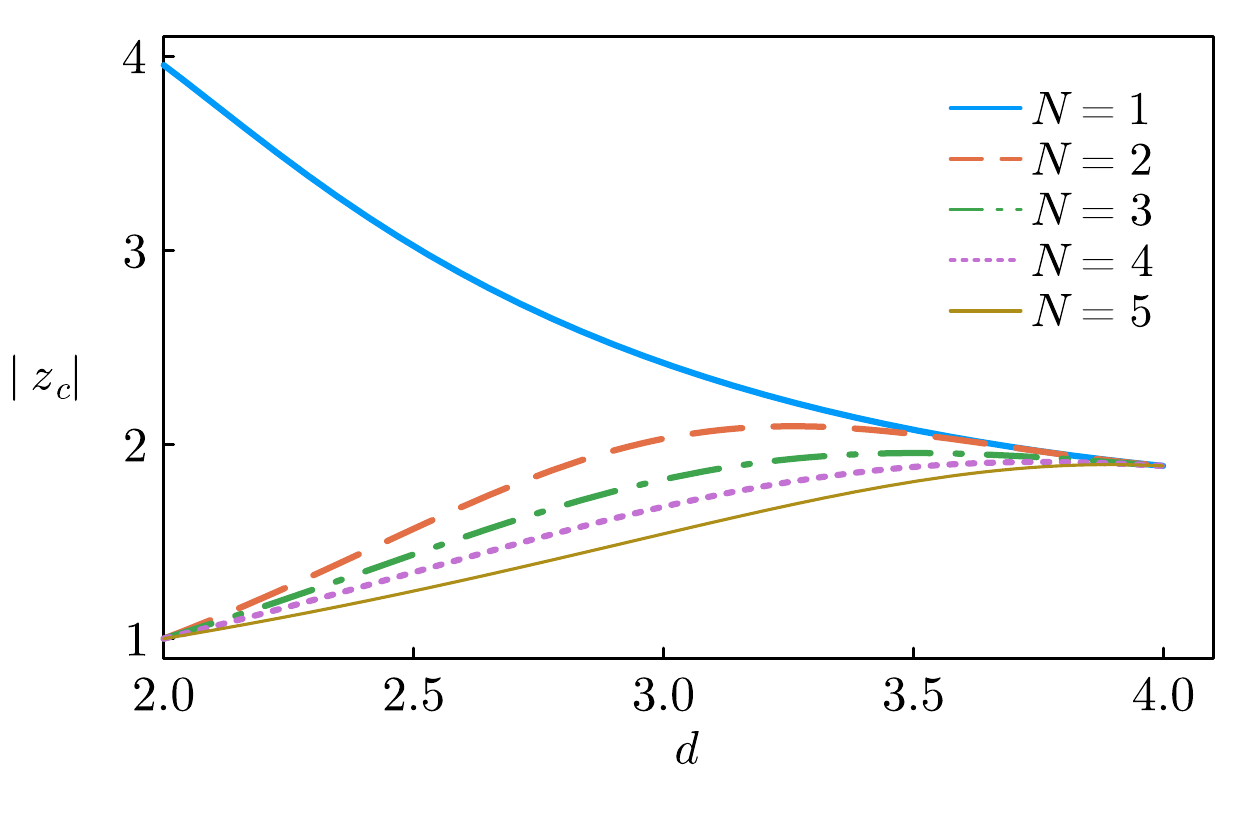}
    \caption{  
    Left panel:  
    the universal location of the \YLE singularity as a function of the number of dimensions $d$ for N=1.   The location of the \YLE singularity in the two-dimensional Ising model (diamond) is obtained from the result of Ref.~\cite{Xu:2022mmw}. 
    The locations in one- and four-dimensions are known analytically. 
    See Ref.~\cite{Rennecke:2022ohx} for details.  
    Right panel: 
four-parameter Pad\'e approximation for the dependence of the \YLE location on the number of spatial dimensions for different numbers of field components, $N$; see~Ref.~\cite{Johnson:2022cqv} for details. 
    }
    \label{fig:N1}
\end{figure}

The flow equation for the wave function renormalization of the longitudinal mode reads  
\begin{align*}
    \notag 
     \partial_t Z_{\|}(\phi) &=
     \int \bar d^d q \partial_t R_k(q^2) \Bigg\{
        G_\|^2 \Big[ \gamma_\|^2 \big(G_\|' + 2 G_\|'' \frac{q^2}{d}\big) 
        + 2 \gamma_\| Z_\|'(\phi) \big(G_\| + 2 G_\|' \frac{q^2}{d}\big)  
         \notag \\ 
        & + (Z_\|'(\phi))^2 G_\| \frac{q^2}{d}
        - \frac12 Z''_\|(\phi) 
        \Big]\notag  \\&
        + (N-1) G^2_{\perp} \Big[ \gamma_\perp^2 \big(G_\perp' + 2 G_\perp'' \frac{q^2}{d}\big) 
        + 4 \gamma_\perp Z_\perp'(\phi)  G_\perp' \frac{q^2}{d}  
          + (Z_\perp'(\phi))^2 G_\perp \frac{q^2}{d}
        \notag \\ &
        + 2 \frac{Z_\|(\phi)-Z_\perp(\phi)}{\phi} \gamma_\perp G_\perp
        - \frac12 \left(\frac{1}{\phi}Z'_\|(\phi)
         - \frac{2}{\phi^2} (Z_\|-Z_\perp)
        \right)
        \Big] 
     \Bigg\}\,.
 \end{align*}
A similar equation for the transverse mode can be found in Ref.~\cite{Johnson:2022cqv}. 
In the above set of equations we used the shorthand notations for 
\begin{align*}
    \gamma_\| &= q^2 Z_\|'(\phi) + U^{(3)}(\phi),  \quad \gamma_\perp = q^2 Z_\perp'(\phi) + \frac{\partial}{\partial \phi} \left( \frac{1}{\phi}U' (\phi) \right)\,, \quad G' = \frac{\partial G} {\partial q^2}. 
\end{align*}

In summary, the flow equation leads to three coupled partial-differential equations on $U$, $Z_\|$ and $Z_\perp$. The equations are stiff and solving them is rather challenging. Instead, one considers Taylor series expansion near $k$-dependent expansion point $\phi_k$ and corresponding equations for the expansion coefficients and $\phi_k$. Traditionally the expansion point is selected to coincide with the minimum of the effective potential $U'_k[\phi_k] = h = $ const. To locate the \YLE, this is however is not the best choice --  it is better to expand near $\phi_k$ defined by   $U_k''[\phi_k] = m^2 \to 0 $. Then, the FRG equation tracks the critical manifold in the parameter space and thus interpolates between the Wilson-Fisher fixed point and the \YLE singularity as illustrated in Fig.~\ref{fig:etaevo}. In the figure, I show the evolution of the effective anomalous dimensions~\footnote{The effective anomalous dimension assumes the value of the actual anomalous dimension at fixed points.}. Without going into details, one can consider the effective anomalous dimension to be a measure of the importance of fluctuations. When fluctuations are not important  $\eta =0$. The dashed line ($d=4$) in this figure demonstrates the fact we considered before: near the critical fixed point, $\eta = 0$ , but as we approach Yang-Lee edge fixed point
the magnitude of $\eta$ increases and reaches a quite large value $\approx 0.3$. This is a manifestation of the fact that the upper critical dimension of the \YLE is larger than the one of the conventional critical point.     

The results are presented in Table~\ref{Tab:zetac}. Note that our approach gives direct access to the universal ratio $|z_c|/R_\chi^{1/\gamma}$ not $|z_c|$. To find $|z_c|$, we used known results for the universal amplitude ratio $R_\chi$ and the critical exponent $\gamma$. Remarkably, FRG calculations can be performed for a non-integer number of dimensions, see Fig.~\ref{fig:N1}. The left panel of the figure demonstrates that 
$|z_c|/R_\chi^{1/\gamma}$ is consistent with results obtained at the linear order of $\varepsilon$-expansion, Eq.~\eqref{eq:zcfromeps} and with two- and  one-dimensional Ising models. The right panel of Fig.~\ref{fig:N1} shows $z_c$ dependence on the number of dimensions and the number of the field components $N$. 
The infinite $N$ limit for the location of the singularity is known exactly~\footnote{The magnetic equation of state in the infinite $N$ limit is given by $f_G(f_G^2 + z)^2 = 1$. An interested reader can easily locate the singularity by finding the solution of the magnetic equation of state and $dz/ df_G =0.$ }. Fif.~\ref{fig:N1} shows
that the approach to this limit is logarithmically slow. 
Reference~\cite{Connelly:2020gwa} demonstrated that $|z_c|$ depends on $N$ non-monotonically and the asymptotic behavior is set in for $N\gg 10$. 

\begin{table}
    \centering
    \begin{tabular}{|c|c|c|c|c|c|}
         \hline
         $N$& 1 & 2 & 3 & 4 & 5 
             \\
         \hline
         \hline
         $|z_c|/R_\chi^{1/\gamma}$& 1.621(4)(1) & 1.612(9)(0) & 1.604(7)(0)  & 1.597(3)(0)     & 1.5925(2)(1) 
         %
         \\ 
 \hline
 $|z_c|$ & 2.43(4) & 2.04(8) & 1.83(6)& 1.69(3) &  1.55(4)\\
         \hline
    \end{tabular}
    \caption{The location of the \YLE singularity, $|z_c|/R_\chi^{1/\gamma}$ and $|z_c|$ for 
     a representative number of components $N$. For $|z_c|/R_\chi^{1/\gamma}$,  the numbers in the parentheses show the truncation error and the error due to residual regulator dependence.  For $|z_c|$, the number in the parentheses represents a combined uncertainty, including that originating from the universal amplitude ratio $R_\chi$ and the critical exponent $\gamma$.  See Ref.~\cite{Johnson:2022cqv} for more details. }
    \label{Tab:zetac}
\end{table}

We established the essential part of the analytical structure near a phase transition, but to have a complete picture, we also need to investigate the symmetries of the QCD partition function. 
We turn to this next. 

\section{QCD analytic structure}
The transition from hadronic gas to quark-gluon plasma along the finite $T$ axis and zero chemical potential was shown to be a smooth ``crossover''. Model calculations, functional methods, and recent analysis of lattice results (described below) hint on the existence of the critical point at some nonzero value of the baryon chemical potential. The crossover changes to a first-order phase transition at the critical point. There is no rigorously defined order parameter to describe the transition. However, due to the smallest of the light quark masses, the chiral condensate is the closest proxy for the order parameter. From the previous discussion, it should be clear that the critical point comes with the corresponding \YLE singularities. To distinguish them from other Yang-Lee edges potentially present in QCD, I will refer to them 
as critical  end point \YLE singularities. 

\subsection{Roberge-Weiss periodicity}
In the introduction, we discussed the properties of the free fermion distribution function including periodicity and singularities in the complex plane of the chemical potential. 
It is necessary to generalize these results to QCD, where the situation is similar qualitatively but quantitatively different. To proceed, consider the QCD partition function at the purely imaginary chemical potential 
\begin{align}
Z (T, \mu = i \theta T ) = {\rm Tr} e ^ {-\frac{\hat H}{T} + i \theta \hat N}\,.  
\end{align}
Due to particle-antiparticle symmetry
$Z(T, -\theta) = Z(T, \theta)$. The partition function can be rewritten as a functional integral~\cite{Kapusta:2023eix}:
\begin{align}
  Z (T, \theta ) &= \int D q  D \bar{q}  D A e^{-S_{E}(T, \theta) }\, , \\
  S_{E}&=\int_{0}^{1/T} d \tau  \int d^{3}x \left(   \bar{q}(\gamma^\mu D_\mu -m) q-\frac{1}{4} G^{2} - i \theta T q^\dagger q \right)\,.
\end{align}
The boundary conditions for the fermionic fields are anti-periodic in the Euclidean time direction $q(x, \tau = 1/T) = - q(x,0)$ and periodic for the gluon field.

The explicit dependence of the Euclidean action on the imaginary chemical potential can be removed by performing the change of variables 
$q(x, \tau) \to e^{i \theta T \tau} q(x,\tau)$. Although the boundary conditions are not invariant under the variable change 
\begin{align}
  q(x, \tau = 1/T) = - e^{i \theta} q(x, \tau=0)\,, 
  \label{eq:q}
\end{align}
this form of the partition function enables us to learn more about QCD symmetries. 

Additionally, consider an aperiodic  gauge transformation 
\begin{align}
  q(x,\tau) &\to U(x,\tau)  q(x,\tau), \\ 
  A(x,\tau) &\to U(x,\tau)  A(x,\tau)  U^{-1}(x,\tau)  - \frac{i}{g} \partial_\mu U(x,\tau)  U^{-1}(x,\tau) . 
\end{align}
Let's choose  $U$ such as to satisfy the boundary condition
\begin{align}
  U(x, \tau = 1/T) = 
  e^{2\pi i k/N_c} U (x, \tau=0)\,, k = 1, \ldots, N_c\,.
  \label{eq:u}
\end{align}
This special class of aperiodic gauge transform is symmetric up to elements of $Z(N_c)$. 

As with any gauge transformation, the action is invariant under~Eq.\eqref{eq:u}. Moreover, the boundary conditions for the gluon field are invariant under the transformation 
due to a general property: the adjoint representation is invariant under the center of the group ($Z(N_c)$ in this case). 
In contrast to the gluon field, the quark field boundary conditions change: 
\begin{align}
  q(x, \tau = 1/T) = - e^{i\left(\theta + \frac{2\pi}{N_c}k  \right)} q(x, \tau =0)
  \label{eq:qZ}
\end{align}
Comparing Eqs.~\eqref{eq:q} and \eqref{eq:qZ}, we conclude that 
\begin{align}
  Z(T, \theta) = Z\left(T, \theta +  \frac{2\pi}{N_c}k \right)\,. 
\end{align}
This is the so-called Roberge-Weiss symmetry. For three colors, we thus have 
the symmetry $\theta \to \theta + 2\pi /3$ or more generally $\hat \mu \to \hat \mu + 2\pi i /3$. Note that this is different from the 
symmetry of the free fermion partition function:  $\hat \mu \to \hat \mu + 2\pi i$.

\begin{figure}
    \centering
       \includegraphics[width=0.29\linewidth]{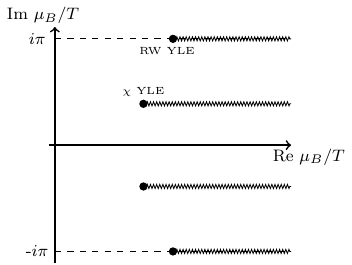}
       \includegraphics[width=0.29\linewidth]{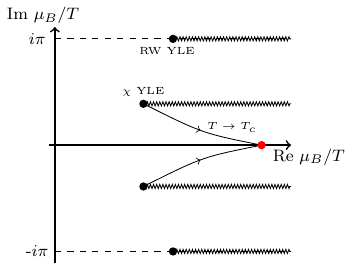}
       \includegraphics[width=0.29\linewidth]{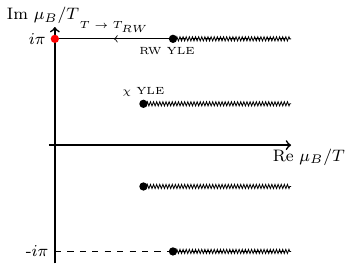}
       \caption{
         An illustration of the complex plane of the baryon chemical potential and the expected location of the Yang-Lee edges associated with the critical endpoint and Roberge-Weiss transition. 
         The panel on the left corresponds to the temperature in the range $ T_c <T < T_{\rm RW}$. When the temperature is lowered  $ T \to T_c$, the critical endpoint \YLE singularities approach and eventually pinch the real axis at $T=T_c$, as 
         demonstrated by the middle panel. The right panel illustrates the case when temperature is increased   $ T \to  T_{\rm RW}$. Again, two corresponding \YLE singularities (note that the figure is symmetric under $\mu_B \to -\mu_B$ approach and pinch imaginary axis. ) 
    }
    \label{fig:RWYLE}
\end{figure}

Similar to free fermions, we expect to have a singularity (we will refer to it as the Roberge-Weiss \YLE) somewhere on the line $\hat \mu = \hat \mu_r + \frac{i \pi}{3}$,
as illustrated in Fig.~\ref{fig:RWYLE}. 

\subsection{Complex chemical potential plane}
Now we are ready to assemble a complete albeit schematic picture of the complex baryon chemical potential plane (see also Ref.~\cite{Stephanov:2006dn}). 
For temperatures in the range $T\in [T_c, T_{\rm RW}]$, we expect to have 
\begin{itemize}
  \item The Roberge-Weiss \YLE singularities on the lines ${\rm Im} \ \hat \mu_B = 3 \ {\rm Im} \hat \mu = \pm i \pi$ ($\pm$ is due to the Roberge-Weiss periodicity)
  \item The \YLE singularities associated with the critical endpoint are located in the complex plane.   
\end{itemize}
We also need to remember that generically the complex plane has the reflection symmetry about both ${\rm Im} \hat \mu_B$ and ${\rm Re} \hat \mu_B$ axes.  
What happens to the position of the critical end point \YLE when the temperature is lowered towards $T_c$? The singularity approaches the real chemical potential axis, and at $T_c$, two critical end point \YLE singularities pinch the real axis. 
This is illustrated by the middle panel in Fig.~\ref{fig:RWYLE}. Now consider the  Roberge-Weiss \YLE. As the temperature is increased towards $T_{\rm RW}$, the singularity slides along the line ${\rm Im} \ \hat \mu_B = \pm i \pi$ 
and at $T_{\rm RW}$, two Roberge-Weiss \YLE singularities pinch the imaginary chemical potential axis (note that $\mu_B= 3 \mu$, as $\mu$ denotes the quark chemical potential).
This is illustrated in the right panel of 
Fig.~\ref{fig:RWYLE}
(only one singularity is shown in the figure). 

Lattice QCD calculations indicate that at $T=T_{\rm RW}$, the Roberge-Weiss critical point coincides with critical point probed by the chiral condensate.
This means that the critical endpoint \YLE either approaches the Roberge-Weiss singularity as $T \to T_{\rm RW}$ or 
joins it at some temperature lower than $T_{\rm RW}$ (neither scenario is shown in the figure). It is currently unknown, and there is no compelling argument supporting either option.

\section{Review of the results based on Lattice calculations}
Lattice QCD simulations do not allow us to locate the \YLE singularities directly due to the sign problem. 
However, one can estimate their location using analytic continuation (e.g., via Pad\'e approximation).
Assuming that the location of the \YLE is found, next, we can analyze its trajectory as a function of temperature and potentially establish the existence/location of the corresponding critical point. 
As far as the critical endpoint \YLE is concerned, an additional complication arises from the lack of high-precision lattice data at low temperatures, where we expect the critical point to reside. 
Thus, one must use the \YLE scaling based on Eq.~\eqref{eq:hc} and extrapolate the \YLE trajectory towards the real baryon chemical potential. Below, we detailed this strategy. However, it is important to validate the strategy for the case when the location of the \YLE is known. This is why I start with the Roberge-Weiss \YLE.    

\subsection{Roberge-Weiss Yang-Lee edge}
\label{Sec:RWYLE}
The key question is if, by tracking the corresponding YLE singularities, one can locate
the position of the critical point in actual lattice QCD calculations. 
In application to Roberge-Weiss critical point, this question is of no direct interest, 
as one can find the critical temperature
by analyzing lattice data at imaginary baryon chemical potential. However, testing the
methods to locate and track the Roberge-Weiss singularity  
constitutes a robust validation of the approach.

At $T<T_{\rm RW}$, the Roberge-Weiss \YLE is located at chemical potential with both non-zero real and imaginary parts. Thus direct determination of the location 
of the singularity in QCD is not possible. The authors of Ref.~\cite{Dimopoulos:2021vrk,Schmidt:2024edp} perform the calculations at purely imaginary values of the baryon chemical potential 
and then analytically continue the results into the complex plane using the Pad\'e  approximation for the first and second derivatives of pressure ($p/T^4$) with respect to the chemical potential ($\hat \mu_B$). For the first derivative, the Pad\`e approximant was taken in the following form    
\begin{equation}
  \chi_1^B(\hat \mu_B)=\frac{\sum_{i=0}^ma_i\hat \mu_B^i}{1+\sum_{j=1}^n b_j\hat \mu_B^j}\;.
\end{equation}
Since the left-hand side of this equation can be computed on the lattice for different imaginary values of $\hat \mu_B$, determining the  
coefficients $a_i,b_i$ simply require solving a set of linear differential equations for each temperature value.
Once the coefficients are known, the zeroes of the denominator can be used to determine the closest singularity $\hat\mu_{\rm YLE}$. It approximates the position of the \YLE 
in the infinite volume limit.

\begin{figure}
    \centering
    \includegraphics[width=0.39\textwidth]{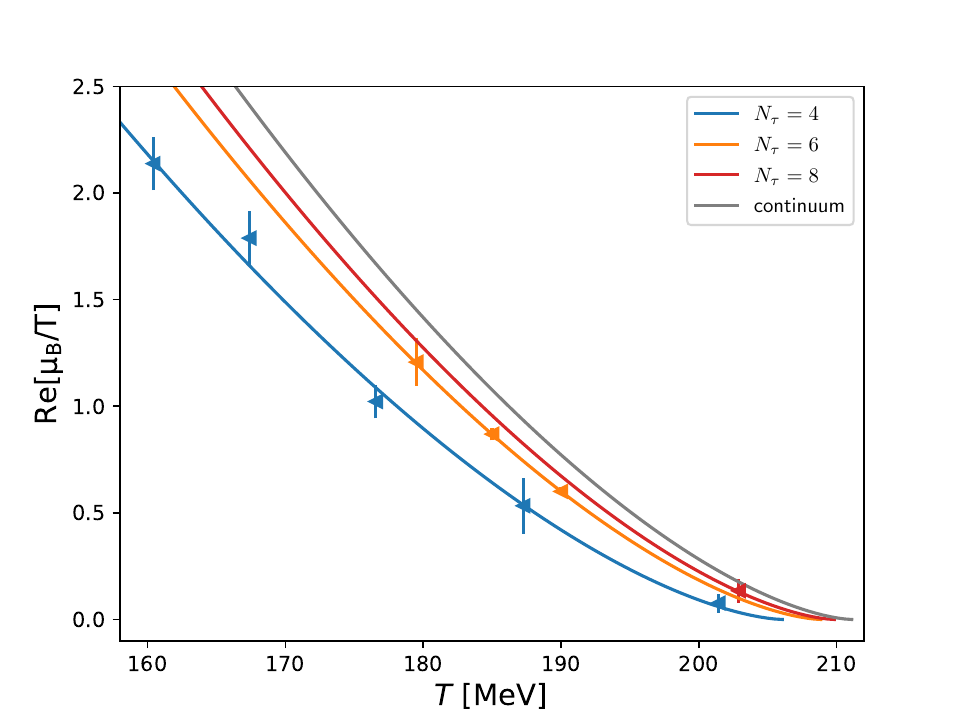}
    \includegraphics[width=0.39\textwidth]{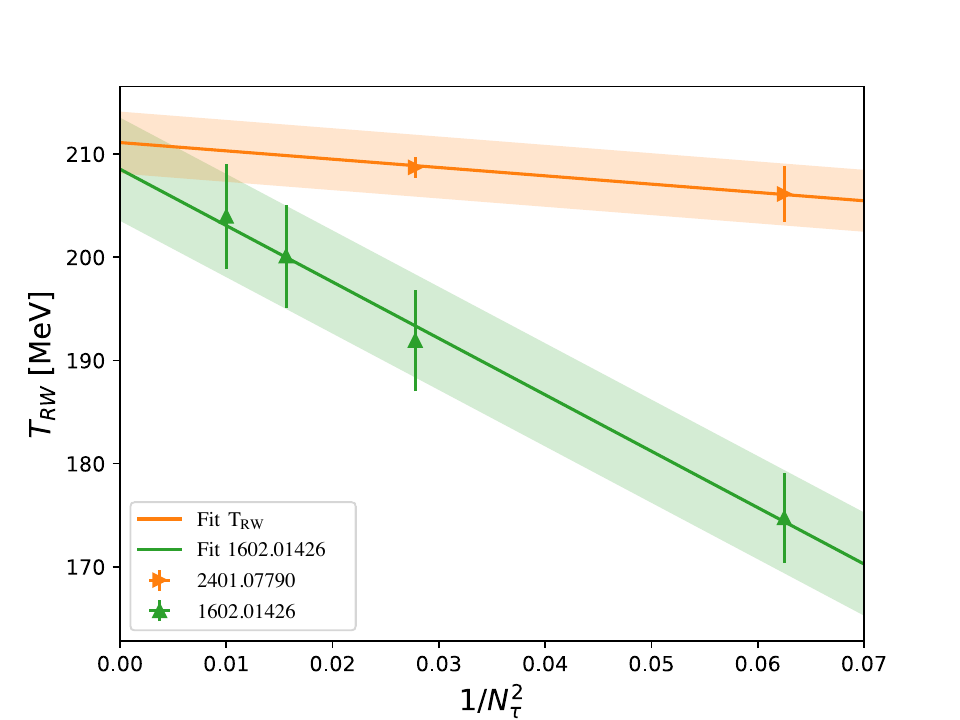}
    \caption{Left panel: scaling of the Yang-Lee singularity with temperature for three different lattice spacings ($N_\tau$) and continuum extrapolation, see     Refs.~\cite{Dimopoulos:2021vrk,Schmidt:2024edp} for details. Right panel: the $N_\tau$ dependence of the Roberge-Weiss transition temperature $\TRW(N_\tau)$ from  Refs.~\cite{Dimopoulos:2021vrk,Schmidt:2024edp} and direct calculation of Ref.~\cite{Bonati:2016pwz}. }
    \label{fig:TRW}
\end{figure}

The next step is to analyze the trajectory of this singularity as a function of temperature. Equation~\eqref{eq:hc}, Yang-Lee scaling, provides a framework for this.  
To make use of it,  $t$ and $h$ have to be expressed 
in terms of the QCD thermodynamic parameters $T$ and $\mu_B$. 
In the vicinity of the Roberge-Weiss critical point (for which the critical chemical potential is known $\hat\mu_B=i\pi$), we have the following mapping relations 
\begin{equation}
  \frac{t}{t_0} = \frac{\TRW-T}{\TRW}, \qquad \frac{h}{h_0}=\frac{\hat\mu_B-i\pi}{i\pi},
\end{equation}
where $t_0, h_0$ are non-universal metric constants and $\TRW$ is the Roberge-Weiss phase transition temperature. 
From Eq.~~\eqref{eq:hc},  we thus have  
\begin{equation}
    \hat\mu_{\rm RW\ YLE}(T)=a \left(\frac{\TRW-T}{\TRW}\right)^{\beta\delta},
    \label{eq:rwlat}
\end{equation}
where the parameter $a$ is a combination of $t_0$, $h_0$ and $|z_c|$. It is to be determined by the fit alongside $\TRW$. 
There is one subtle point to be taken into account before one  determines $\TRW$: lattice cut-off dependence or the 
dependence on the $N_\tau$. By fitting parameters, $a$ and $\TRW$ in equation~\eqref{eq:rwlat} one finds them at a given value of $N_\tau$. In Ref.~\cite{Schmidt:2024edp}, it was assumed that the 
cut-off dependence of the Roberge-Weiss critical temperature and the amplitude is given by 
\begin{align}
  \TRW(N_\tau)&=\TRW^{(0)}+\TRW^{(2)}/N_\tau^2, \\
  a(N_\tau)&=a^{(0)}+a^{(2)}/N_\tau^2 ,
\end{align}
which is justified for large enough $N_\tau$.  
All together, one ends up with four fitting parameters.  The left panel of Fig.~\ref{fig:TRW}
demonstrates the quality of the scaling~\eqref{eq:rwlat} at a given value of $N_\tau$, while 
the right panel shows the extrapolation of $\TRW$ to the continuum limit,  $\TRW^{(0)}=211.1\pm 3.1$ MeV.  
Comparison with the direct calculations at the imaginary value of the chemical potential from Ref.~\cite{Bonati:2016pwz} shows a good agreement with  the 
result obtained by tracing the \YLE, see the right panel of  Fig.~\ref{fig:TRW}.

This serves as a proof of principle for locating the critical point by tracing the location of the \YLE. 
Can a similar strategy be executed for a QCD critical point at a finite real chemical potential?  The answer is yes, as the recent articles \cite{Basar:2023nkp,Clarke:2024ugt} demonstrated. I will now summarize their methods and results.

\subsection{\YLE associated with critical end  point}

\begin{figure}
    \centering
    \includegraphics[width=0.45\linewidth]{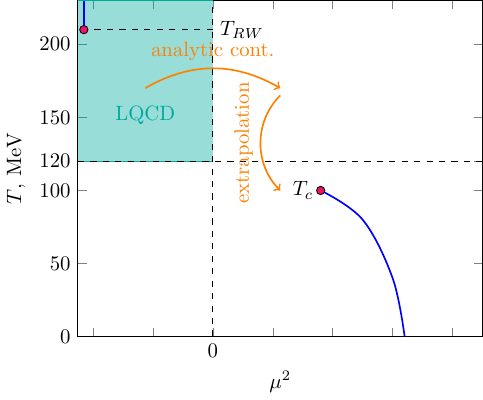}
    \caption{The figure illustrates the domain of temperature and chemical potential where Lattice QCD performed direct simulations: imaginary (negative $\mu^2$) and zero baryon chemical potential and at temperatures approximately above 120 MeV. To access thermodynamics at real values of the chemical potential (positive values of $\mu^2$), an analytic continuation has to be performed. It has to be supplemented by some physics-motivated extrapolation technique to access temperatures lower than 120 MeV.  }
    \label{fig:qcdpd}
\end{figure}

As I alluded to before, the main challenge of locating the QCD critical point with the current lattice data can be attributed to 
the lack of the low-uncertainty low-temperature calculations. For example, the current usable data at imaginary chemical potential reaches $T\approx 120$ MeV. At the same time, there is a strong indication from the functional methods that the critical point is at about 100 MeV. Thus, the \YLE trajectory cannot be found near the possible critical point, and one has to rely on the extrapolation of the \YLE trajectory to lower temperatures. Together with the statistical uncertainty of the lattice data and the systematic error of the analytical continuation, the extrapolations constitute the major sources of uncertainty for determining the location of the critical point. 

I describe the extrapolation of the trajectory first, as it is common for both strategies described below. 
It is similar in idea to the analysis of the Roberge-Weiss singularity but different in the actual realization.
Assume we could locate the \YLE at a set of temperatures $T>T_c$. This input allows us to fix unknown 
non-universal numbers in the relation~\eqref{eq:hc}. For this, we need to establish the relationship between 
$t$ and $h$, the actual temperature, and the chemical potential. In the case of the critical point, we have a mixture 
of the reduced temperature $\Delta T = T-T_c$  and the reduced chemical potential $\Delta \mu = \mu-\mu_c$  contributing to the scaling variables 
\begin{align}
  h & = a \Delta T + b \Delta \mu, \\ 
  t & = c \Delta T + d \Delta \mu\,.  
\end{align}
Substituting to Eq.~\eqref{eq:hc} we have 
\begin{align}
  a \Delta T + b \Delta \mu  = i |z_c|^{\beta \delta} (c \Delta T + d\Delta \mu)^{\beta\delta}\,. 
  \label{Eq:CPsc}
\end{align}
Considering small $\Delta T$  and  $\Delta \mu$ and taking into account that for Z(2) universality class 
$\beta \delta \approx 3/2$ $> 1$  
at the leading order, we have
\begin{align*}
  a \Delta T + b \Delta \mu = 0
\end{align*}
or 
\begin{align}
  {\rm Re}\, \Delta \mu_{\rm LO} = - \frac{a}{b} \Delta T \,. 
  \label{Eq:LOREmu}
\end{align}
Thus, the real part of the reduced chemical potential approaches zero linearly as a function of temperature. 
The imaginary part of the reduced chemical potential can be easily computed by substituting Eq.~\eqref{Eq:LOREmu}
 to  \eqref{Eq:CPsc} and taking the imaginary part. The leading imaginary part is then 
\begin{align}
  {\rm Im}\, \Delta \mu_{\rm LO} =  \frac{1}{b} |z_c|^{\beta \delta} \left(c - \frac{da}{b}\right)^{\beta \delta}  \Delta T^{\beta\delta} \,. 
  \label{Eq:LOIMmu}
\end{align}
That is, the imaginary part of the reduced chemical potential approaches zero faster than the real part. Of course, there will be higher-order corrections for both equations. The data analysis suggests that one has to add $\Delta T^2$ to the real part to fit $ {\rm Re}\, \Delta \mu$ as a function of temperature.  
For the sake of simplicity, it is convenient to lump together different combinations of constants into new  coefficients in front of different powers of $\Delta T$: 
\begin{align}
  {\rm Re}\, \mu = \mu_c + c_1 \Delta T + c_2 \Delta T^2 , \\ 
  {\rm Im} \mu = c_3 \Delta T^{\beta \delta} .
  \label{Eq:ReImCh}
\end{align}
All in all, there are five unknown coefficients $c_1$, $c_2$, $c_3$, $T_c$ and $\mu_c$. They have to be fixed using the lattice data.   

\subsubsection{Purely imaginary chemical potential data}
\begin{figure}[t]
  \begin{center}
    \includegraphics[height=0.29\textwidth]{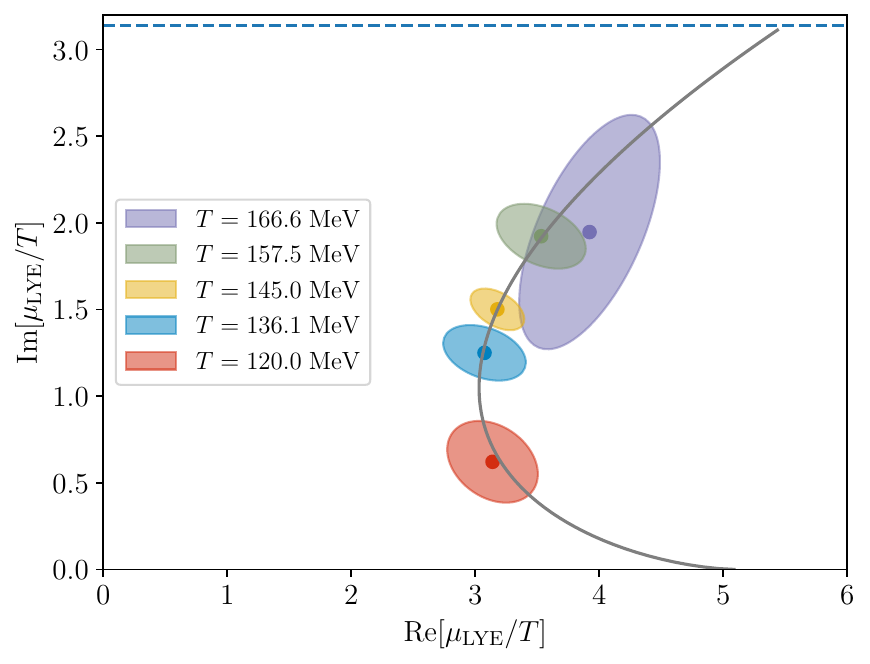}
    \includegraphics[height=0.29\textwidth]{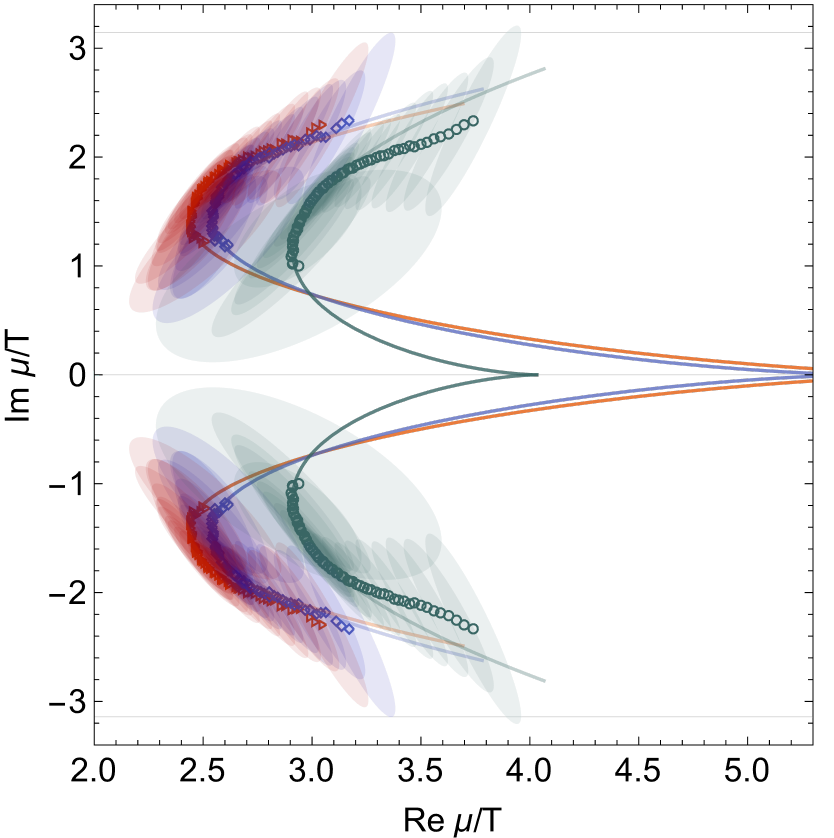}
  \end{center}
  \caption{
  Left panel: the figure is from Ref.~\cite{Clarke:2024ugt}. It shows the location of the \YLE singularity as a function of temperature for $N_\tau =6$  extracted using the imaginary baryon chemical potential data. The solid line is the result of the best fit with Eq.~\eqref{Eq:ReImCh}. 
  Right panel: the figure is from Ref.~\cite{Basar:2023nkp}. In the figure, the location of the \YLE singularity is displayed as a function of temperature extracted from Taylor series coefficients for $N_\tau=8$. The \YLE locations (data points) and corresponding fits (solid lines) correspond to different extraction methods.  The red triangles and the blue diamonds are obtained by utilizing two different conformal maps with subsequent Pad\'e approximation. The green circles are obtained by using Pad\'e without any conformal map.    
  }\label{fig:traj}
\end{figure}

The general strategy is the same as the one described in Sec.~\ref{Sec:RWYLE}. That is, the lattice data is collected at imaginary values of the baryon chemical potential and then analytically continued into the complex plane using multi-point Pad\'e. The closest zero of the denominator in Pad\'e approximant estimates the position of the \YLE. Of course, the data must be collected at lower temperatures, and the fit must be performed with Eq.~\eqref{Eq:ReImCh}. 
The fit of the imaginary $\mu_B$ data was done within the interval $[\hat \mu_{B {\rm min}}, \hat \mu_{B {\rm max}} ] \in [-i\pi, i\pi]$. The length of the interval $|\hat \mu_{B {\rm min }}  - \hat \mu_{ B {\rm max }}|$ was varied between $\pi$ and $2\pi$. Overall, the authors of Ref.~\cite{Clarke:2024ugt} perform 55 Pad\'e approximations per one temperature value. Only zeroes of the denominator  closest to $[\hat \mu_{B {\rm min}}, \hat \mu_{B {\rm max}} ]$ were used. This provides a handle on the systematic 
uncertainty of the extraction of the \YLE singularity. 
I refer the reader to Ref.~\cite{Clarke:2024ugt} for more technical details.
The results are presented in Fig.~\ref{fig:traj}. 
The data was analyzed for $N_\tau=6$ and 8. The largest $N_\tau$ yields the location of the critical point at $T_c = 101\pm15 $ MeV, $\mu_c = $ 560$\pm140$ MeV. As the authors point out, the continuum limit will most likely result in larger chemical potential values of about 650 MeV~\cite{Clarke:2024ugt} and slightly higher temperatures of about 110 MeV  \cite{CPOD24}. 
  \begin{figure}[b]
  \begin{center}
    \includegraphics[height=0.2\textwidth]{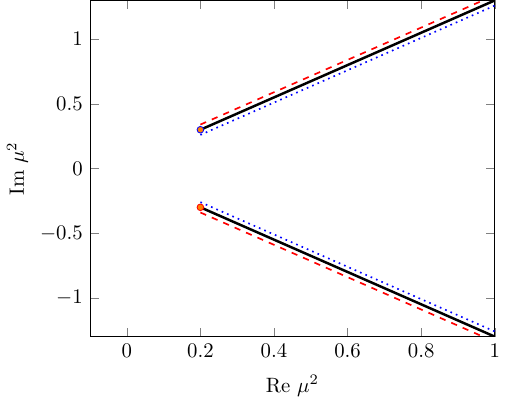}
    \includegraphics[height=0.2\textwidth]{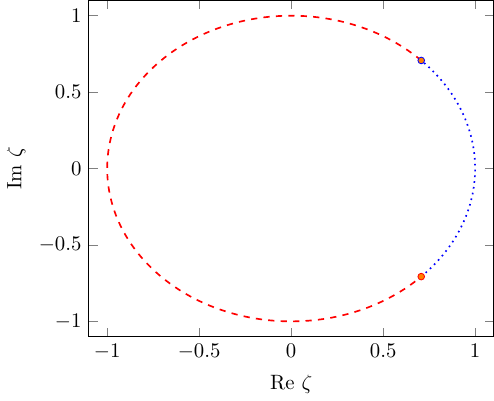}
  \end{center}
  \caption{Illustration of the ``two-cut'' map of Ref.~\cite{Basar:2023nkp}. 
  }\label{fig:mapping}
\end{figure}

  \subsubsection{Taylor series data}
  Analysis of Ref.~\cite{Basar:2023nkp} is similar, but there are the following key differences
  \begin{itemize}
    \item Taylor series expansion coefficients computed at zero chemical potential were used. Note that currently, only eight coefficients are known. The  highest two of them lack the continuum limit estimate and are computed at a fixed $N_\tau = 8$.
    \item Application of the Pad\'e analysis is done after  the conformal mapping applied  to help improve the extraction of singularity. That is, first, one performs the change of the variable $\mu^2 = f(\zeta)$, next obtains the Taylor series in $\zeta$ about $\zeta_0$ defined by $f(\zeta_0) = 0$, and finally performs Pad\'e in $\zeta$ variable.   The reason to perform a conformal map before Pad\'e analysis is that when the function is expected to have a complex conjugate pair of singularities, Pad\'e results in two arcs of poles that emerge from singularities; they coalesce along the real axis. 
These poles limit the radius of convergence of the Pad\'e approximant. The idea is to map them beyond the disk of the radius defined by the \YLE.  
  \end{itemize}

The function $f$ is chosen such as to map the YLE cuts onto the unit circle. A figure best demonstrates this; see Fig.~\ref{fig:mapping}. The red and blue lines are mapped onto the unit circle. The function that performs this ``two-cut'' transformation is   
\begin{equation}
f(\zeta) = 4\left|\mu_{\rm YLE}^2\right| \zeta\left[\frac{\theta / \pi}{(1-\zeta)^2}\right]^{\theta / \pi}\left[\frac{1-\theta / \pi}{(1+\zeta)^2}\right]^{1-\theta / \pi}\,.
\end{equation}
This transformation has to be supplied by $\mu_{\rm YLE} = |\mu_{\rm YLE}| e^{i \theta}$.
This is, however, exactly what we want to find. The author of Ref.~\cite{Basar:2023nkp} adopted the following strategy: first, a rough estimate for the  \YLE location is found using the result of pure Pad\'e analysis, then it is improved by 
applying the map, expanding in $\zeta$ and performing the Pad\'e analysis of the corresponding Taylor series in $\zeta$. 
This procedure is then iterated until it converges to a fixed point that defines  $\mu_{B {\rm YLE }}$.  
Beyond this, the uniformizing map was performed in Ref.~\cite{Basar:2023nkp}. However, the result is very close to the ``two-cut'' map and thus will not be described here.   

In Fig.~\ref{fig:traj}, the result of this procedure is shown. Blue diamonds show the two-cut map, while the pure Pad\'e gives green circles.

\subsubsection*{An example of improving Taylor series convergence with a conformal map}
\label{Sec:Conformal}
It seems that the first step (conformal map)  plays a crucial role, and to demonstrate one of the advantages of performing a conformal map for the convergence of the Taylor series, let us return to the example we considered in Introduction, Eq.~\eqref{Eq:TaylorSample}.  
We will perform the following set of conformal maps. Firstly, we map $\mu$ to the fugacity plane 
$$\lambda = e^{\hat \mu}.$$
\begin{figure}
  \begin{center}
    \includegraphics[height=0.2\textwidth]{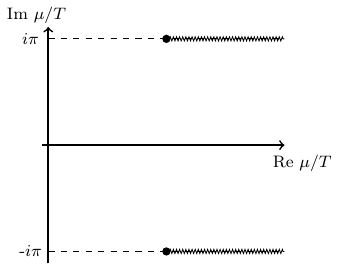}
    \includegraphics[height=0.2\textwidth]{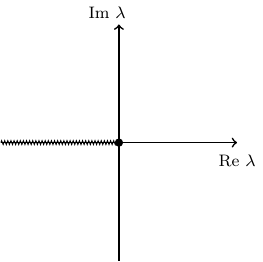}
    \includegraphics[height=0.2\textwidth]{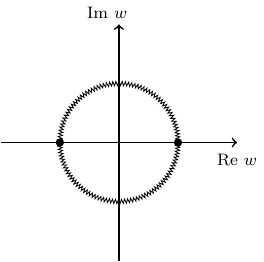}
  \end{center}
  \caption{Series of maps Eq.~\eqref{Eq:mapw} and Eq.~\eqref{Eq:maplamb}.
  As a result, the strip $-\pi<$ Im $\hat \mu<\pi$ is mapped onto the unit disk $|w|<1$.      
  }
  \label{fig:mappingF}
\end{figure}
This map accounts for the periodicity of our function and transforms the lines $\hat \mu = \pm i \pi$ to the negative real axis. Secondly, we map the negative real axis onto  the circumference of the unit
circle using 
\begin{align}
w = \frac{\sqrt{\lambda} -1} {\sqrt{\lambda} +1} 
\label{Eq:mapw}
\end{align}
with the inverse 
\begin{align}
\lambda = \frac{(w+1)^2}{(w-1)^2}\,.
\label{Eq:maplamb}
\end{align}
The expansion about $\hat \mu = 0$ corresponds to the expansion about $w = 0$. For our example, we set $\hat \omega =1$ and obtained the radius of convergence in 
$\hat \mu$ to be $R_{\hat \mu} = \sqrt{1 + \pi^2}$. In the variable $w$, the radius of convergence is $R_{\omega} = 1$.
That is, after mapping, the function can be extracted to values of $\hat \mu$: 
\begin{align*}
\left| \frac{{e^{\hat \mu/2}} -1}{{e^{\hat \mu/2}} + 1}\right|  < 1 
\end{align*}
or simply 
\begin{align*}
\left| {e^{\hat \mu/2}} -1 \right|  <   \left|  {e^{\hat \mu/2}} + 1\right|  
\end{align*}
which is true for any $\hat \mu$, thus allowing us to recover our function from Taylor series expansion for any $\hat \mu$. By doing the maps (see illustration in Fig.~\ref{fig:mappingF}), we gained a lot: it became possible for us to extend the finite range of convergence
to an unbounded one. We achieved this by incorporating information about the analytic structure of our function into the map.

\begin{figure}[t]
  \begin{center}
    \includegraphics[width=0.3\textwidth]{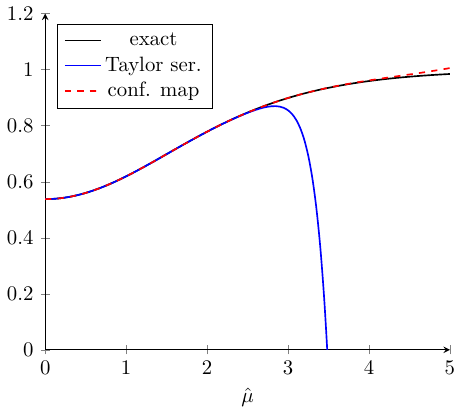}
  \end{center}
  \caption{Illustration of the convergence of the truncated Taylor series expansion in $\bar \mu$  and the one in $w$ after performing the conformal map. Both series are truncated at order $20$. The figure demonstrates how a conformal map 
  can improve the convergence of the Taylor series without providing any extra Taylor series coefficients. }
  \label{fig:mappedres}.
\end{figure}

\begin{figure}[b]
  \begin{center}
    \includegraphics[width=0.3\textwidth]{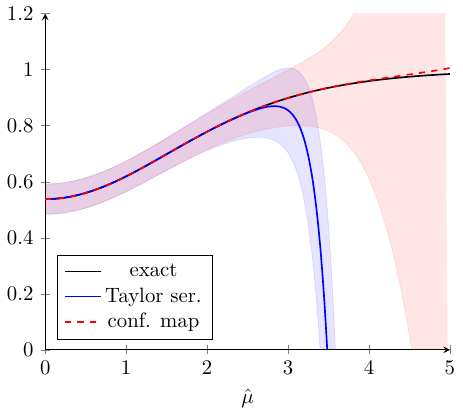}
  \end{center}
  \caption{
  The same as in Fig.~\ref{fig:mappedres}, but with 10\% uncertainty added to the original Taylor series coefficients.  }
  \label{fig:mappedresErr}.
\end{figure}

To demonstrate this explicitly,  start with a truncated series expansion in $\hat \mu$ ($a_i$ is non-zero for even $i$ only): 
\begin{align*}
  f(\hat\mu) = \sum_{i=0}^{M} a_i \mu^{i}   
\end{align*}
Next, substitute $\mu = \ln \frac{(w+1)^2}{(w-1)^2}$ (the inverse of the two maps performed) into this expression and expand in $w$ about zero (due to $\omega(\mu = 0) = 0 $) to the same truncation order:    
\begin{align*}
  f(w) = \sum_{i=0}^{M} b_i \omega^{i} ,  
\end{align*}
where $\omega$ is a function of $\mu$, $\omega = \frac{{e^{\hat \mu/2}} -1}{{e^{\hat \mu/2}} + 1}$. Finally, plot the resulting $f(\mu)$. This procedure is illustrated in Fig.~\ref{fig:mappedres} for $M=20$. 
We see a significant improvement in reproducing the exact function. Of course, there is a nuance.  We assumed that the coefficients $a_i$ are known precisely and that linear combination does not lead to an exploding uncertainty. To study this question, let's consider a few terms in the expansion of $f(w)$ after performing the conformal map and reexpanding the series; we have  
\begin{align*}
  f(w) = a_0 + 16 a_2 w^2 + \left( \frac{32}3 a_2 + 256 a_4 \right) w^4 + \ldots 
\end{align*}
This demonstrates that indeed the coefficients $b_j$ are just linear combinations of $a_{i\leq j}$, specifically 
\begin{align*}
  b_0 &= a_0, \\
  b_2 &= 16 a_2, \\
  b_4 &= \frac{32}{3} + 256 a_4 , 
   \\ & \cdots  
\end{align*}

If $a_i$ are known with some finite precision, one might get concerned about large coefficients appearing in $b_2$ in front of $a_2$ or  $b_4$ in front of $a_2$ and $a_4$. 
However, in this specific case, this would be unfounded. Consider the quadratic term, $a_2\hat \mu^2$ for $\hat \mu = 5$. It equals $25 a_2$. At the same time, taking into account that 
$w(\hat \mu= 5 ) \approx 0.85$ we have $16 a_2 w^2 \approx 12 a_2$. Thus, for the quadratic truncation,  the actual coefficients in front of $a_2$ are not larger for the series expansion after the map. 
However, at higher orders, we have more terms contributing to a $b_j$, and the situation with the uncertainty estimate can drastically change. This is more straightforward to demonstrate with an actual computation. I will continue working with the same function but prescribe a $10\%$ uncertainty to each coefficient $a_i \to a_i  \pm a_i/10$. I also assume there is no correlation between the uncertainties for each coefficient. 
After this, the analysis can be repeated. Its result is shown in Fig.~\ref{fig:mappedresErr}.
The figure demonstrates that most of our gains with the conformal map are wiped away by a 10\% uncertainty in determining the coefficients $a_i$. This example should be taken as a caution tale; it cannot provide a universal conclusion of the usefulness of a conformal map for the case when Taylor series coefficients are known with a given finite precision.       

\section{What else \YLE singularity affects?}
\subsection{Asymptotic Taylor series}

In the previous section, I reviewed that the location of the \YLE singularity can be estimated by analyzing Taylor series coefficients. 
One can go a step further in this direction analytically and demonstrate that the location and the edge critical exponent are imprinted in 
high order Taylor series coefficients  due to Darboux theorem~\cite{henrici1977applied,doi:10.1137/0139022}. To demonstrate this, consider a function $f(x)$  with singularity closest to the origin located at the real point $x=x_1$. Suppose that the behavior of this function near the singularity~\footnote{This form neglects the confluent singularity. } is 
\begin{align}
& f(x)=\left(1-x / x_1\right)^{\sigma+1} r(x)+a(x) 
\label{Eq:fx}
\end{align}
where $r(x)$ is an analytic in some disk centered at $x_1$ and having radius larger than $|x_1|$. 
\begin{align}
& r(x)=\sum_{k=0}^{\infty} b_k\left(x-x_1\right)^k 
\label{Eq:rx}
\end{align}
Similarly, $a(x)$ is assumed to be analytic in some disk with center $x=0$ and radius  $>\left|x_1\right|$. 
The critical exponent  $\sigma$ is non-integer (the case of interest for us). 
The assumed form of the function in Eq.~\eqref{Eq:fx} is quite restrictive but is sufficient for our purposes.

Expanding $\left(1-x / x_1\right)^{k + \sigma+1}$ into a power series about $x=0$, we have 
\begin{align*}
   \left(1-x / x_1\right)^{k+ \sigma+1} = 
   \sum_{n=0}^{\infty} 
   \frac{ \Gamma (n  - k -1 - \sigma)}{\Gamma (n+1) \Gamma (-1-k-\sigma) } \left(\frac{x}{x_1}\right)^{n} \,.
\end{align*}
Combining this equation with Eq.~\eqref{Eq:rx}, we get 
\begin{align*}
   \left(1-x / x_1\right)^{\sigma+1}  r(x) = 
   \sum_{n=0}^{\infty}  \sum_{k=0}^{\infty} (-1)^k b_k x_1^{k-n} 
  \frac{ \Gamma (n  - k -1 - \sigma)}{\Gamma (n+1) \Gamma (-1-k-\sigma) }  x ^{n} \,.
\end{align*}
That is, the asymptotic coefficients of the expansion are given by  
\begin{align}
   c_n = 
  \sum_{k=0}^{\infty} (-1)^k b_k x_1^{k-n} 
  \frac{ \Gamma (n  - k -1 - \sigma)}{\Gamma (n+1) \Gamma (-1-k-\sigma) }   . 
  \label{Eq:cn}
\end{align}
In general, the expansion coefficients of the function $f(x)$ are the sum of $c_n$ and $a_n$ ($a(x) =$ $ \sum a_n x^n$). 
However, the asymptotic expansion coefficients of $f(x)$ are not affected by $a_n$ and are given by $c_n$ only.  
To show this, consider  the closest singularity of $a(x)$ to the expansion point $x=0$ at some $x=x_2$ with  $\left|x_2\right|>\left|x_1\right|$. Compared to $c_n$, $a_n$ are  
smaller  by a factor $O\left(\left|x_1 / x_2\right|^n\right)$, and hence are negligibly small asymptotically.

Applying  Stirling's formula for $n\to\infty$ to the ratio  
\begin{align*}
\frac{ \Gamma (n  - k -1 - \sigma)}{\Gamma (n+1)  }   = 
n^{-k-\sigma -2} \left(1+\frac{(k+\sigma +1) (k+\sigma +2)}{2
   n}+O\left(\left(\frac{1}{n}\right)^2\right)\right)
\end{align*}
demonstrates 
that successive terms in Eq.~\eqref{Eq:cn} are suppressed by inverse powers of $n$. For sufficiently large 
 $n$ we thus have 
\begin{align}  
c_n \sim  b_0 x_1^{-n} \frac{ \Gamma (n  -1 - \sigma)}{\Gamma (n+1) \Gamma (-1-\sigma) }   
\sim 
\frac{ b_0 x_1^{-n} }{n^{2+\sigma} \Gamma (-1-\sigma) }  \,. 
\end{align}
In  QCD, as we discussed above, there are two complex conjugate singularities in the $\hat \mu_B^2$ plane. Both are the same distance from the expansion point (the origin). Therefore, the above analysis
has to be amended. This can be easily done in the same fashion as above with the result: 
\begin{align}
  \notag 
  c_n &\sim 
  \frac{ b_0 x_1^{-n} }{n^{2+\sigma} \Gamma (-1-\sigma) }  + {\rm c.c.} \\ 
    &= \frac{ 2 |b_0| |x_1|^{-n} }{n^{2+\sigma} \Gamma (-1-\sigma) } \cos(\phi_{b_0} - n \phi_{x_1}) ,  
    \label{eq:cnas}
\end{align}
where $\phi_{b_0}$ and $\phi_{x_1}$ are the phases of $b_0$ and $x_1$ respectively. In application to QCD we interpret $x\to \hat \mu_B^2$ and 
$x_1 \to \hat\mu^2_{ B {\rm YLE} }$.  

We thus conclude that the location of the \YLE   and its critical exponent $\sigma$ define the asymptotic behavior of 
Taylor series coefficients. Of course, to gain anything practical from this analysis in QCD, 
one has to compute high-order Taylor series expansion coefficients. As I pointed out before, the highest available Taylor series coefficients from lattice QCD calculations are of a modest eighth order. 
At what order the asymptotics of Eq.~\eqref{eq:cnas} set in is unknown and would require detailed analysis, most likely involving higher order Taylor coefficients.  

\subsection{Fourier coefficients}
Another striking example of the \YLE influence on thermodynamics 
is the Fourier coefficient of the baryon chemical potential. 
\begin{figure}[b]
    \centering
    \includegraphics{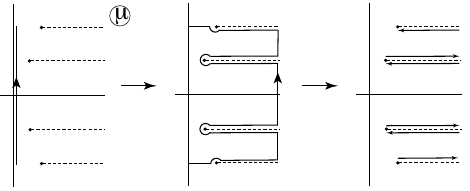}
    \caption{Complex chemical potential plane with the critical end point and RW YLE. The integration path along the imaginary chemical potential axis can be deformed so that the contour follows around the branch point singularities and the cuts. }
    \label{fig:countour}
\end{figure}
As we established in Sec.~\ref{Sec:RWYLE}, 
the QCD partition function is periodic in baryon chemical potential: 
$ Z\left(\hat \mu_B + 2\pi i \right) = Z\left(\hat \mu_B\right) $. 
Due to this periodicity, analysis of the data obtained for purely imaginary values of baryon chemical potential is natural in terms of the corresponding Fourier coefficients~\cite{Roberge:1986mm, Vovchenko:2017xad,Almasi:2018lok, Bornyakov:2016wld,Bornyakov:2022blw, Schmidt:2023jcv}. 
Usually one computes Fourier transformation of the baryon number density $n_B = V^{-1} \partial_{\hat \mu_B} \ln Z(\hat \mu_B)$:
\begin{align}
    \label{Eq:bk}
    b_k = \frac{1}{i \pi} \int_{-\pi}^{\pi} d\theta  \,  \hat n_B (\hat \mu_B = i \theta ) 
    \sin \left(k \theta \right),
\end{align}
where I explicitly took into account the symmetry property of the baryon density $ n_B(\hat \mu_B) = $ $-n_B(-\hat \mu_B)$ and introduced $\hat n_B = n_B/T^3$. 

The Fourier coefficients are sensitive to the structure of the phase diagram through the following line of arguments, see Ref.~\cite{Bryant:2024twd} for details.    Consider the integral~\eqref{Eq:bk} in the complex plane,
see Fig.~\ref{fig:countour} for an illustration. 
 Since the continuous deformation of the contour does not change the value of the integral as long as it does not cross the singularities, one can modify the integration to that over both sides of the cuts, as shown in~Fig.~\ref{fig:countour}. To proceed with an actual realization of this program, it is helpful first to consider just one singularity and the corresponding cut in isolation.

\begin{figure}
    \centering
    \includegraphics{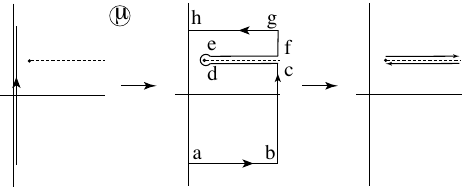}
    \caption{Computation of the Fourier coefficient for the function with one singularity in the right half-plane of complex baryon chemical potential.}
    \label{fig:2}
\end{figure}

Consider an odd  function $n_B (\hat \mu_B)$ periodic in an imaginary argument having  brunch points in the complex plane located at 
$\pm \hat  \mu_B^{\rm br}$, where 
$ \hat   \mu_B^{\rm br} = \hat  \mu_r^{\rm br} + i\, \hat  \mu_i^{\rm br} $.
Here, we expand $n_B$ near the branch point 
  $\hat \mu_B \to + \hat  \mu_B^{\rm br}$:  
\begin{align}
\hat n_B(\hat \mu)  =  A (\hat \mu_B - \hat \mu_B^{\rm br})^{\sigma} 
(1 + B(\hat \mu_B - \hat \mu_B^{\rm br})^{\theta_c} + \ldots )  + 
\sum\limits_{n=0}^{\infty} a_n (\hat \mu_B - \hat \mu_B^{\rm br})^n
\label{Eq:NbExp}
\end{align} with $\sigma>-1$ and $\theta_c>0$. In the context of the YLE singularity, $\theta_c$ is the confluent critical exponent (not to be confused with integration variable $\theta$).   The regular part of $\hat n_B$ on the cuts is encoded by the coefficients $a_n$. 

From the definition of the Fourier coefficients, we have 
\begin{align}
    b_k = \frac{1}{\pi} \int_{-\pi}^{\pi} d\theta  \,  \hat n_B (\hat \mu_B = i \theta ) 
   e^{-i k \theta} \,.
\end{align}
 To compute the integral, we will deform the contour as shown in Fig.~\ref{fig:2}. The figure assumes that the right-most points are extended to infinity, i.e., ${\rm Re} \, \mu \to \infty$. The contribution of the segments $(ab)$ and $(gh)$ cancel each other due to the integrand's periodicity and the segments' opposite direction. 
The contributions from $(bc)$ and $(fg)$ is zero due to the exponential decay of $\exp(- i k \theta  ) 
= \exp(- k \hat \mu_B  ) $ for any $k>0$. The integral around the branch point  $(de)$ vanishes due to $\sigma>-1$. Thus, only the segments on both sides of the cut $(cd)$ and $(ef)$  give a non-trivial contribution to the integral, as illustrated in  Fig.~\ref{fig:2}. 
To evaluate the contribution of these segments, we consider the parametrization 
$i \theta = s  + \hat \mu_B^{\rm br}$
\begin{align}
    \frac{1}{\pi} \int_{(ef)} d\theta  \,  \hat n_B (\hat \mu_B = i \theta ) 
   e^{-i k \theta} 
   &=     \frac{1}{i \pi} e^{-\mu_B^{\rm br} k } \int_{0}^{\infty} d s  \,  \hat n_B (\hat \mu_B = s  + \hat \mu_B^{\rm br} ) e^{- k s} \,.
\end{align}
Evaluating the  integrals, one obtains  
\begin{align}
 \nonumber
    \frac{1}{\pi} \int_{(ef)} d\theta  \,  \hat  n_B (\hat \mu_B = i \theta ) 
   e^{-i k \theta} 
   &=  \frac{ e^{-\mu_B^{\rm br} k }}{i \pi}  \left(
   A \frac{\Gamma(1+\sigma)}{k^{1+\sigma}} \left[ 
    1 +   \frac{B}{k^{\theta_c}}   \frac{\Gamma(1+\sigma + \theta_c)}{\Gamma(1+\sigma)} 
   + \ldots \right]
    \right. \\ &\left. 
    + \sum_{n=0}^\infty a_n \frac{\Gamma(1+n)}{k^{1+n}} 
   \right)     \,.
\end{align}
In the second line, the sum is due to the analytic part in Eq.~\eqref{Eq:NbExp}.

The integral over the segment $(cd)$ is identical to the expression above except for the $2\pi$ rotation around the branch point and an extra minus sign due to the direction of the segment.
Adding both integrals together cancels the analytic part to yield 
\begin{align}
    b_k = 
    \frac{ e^{-\mu_B^{\rm br} k }}{i \pi}  
   A \frac{\Gamma(1+\sigma)}{k^{1+\sigma}} \left(
    1 - e^{i 2\pi\sigma} +   \frac{B}{k^{\theta_c}} \left[ 1 - e^{i 2\pi(\sigma+\theta_c)} \right]   \frac{\Gamma(1+\sigma + \theta_c)}{\Gamma(1+\sigma)} 
   + \ldots \right)\,.
\end{align}
Absorbing $k$ independent factors into constants $A$ and $B$, we arrive at  
\begin{align}
    b_k = 
      \tilde A \frac{e^{-\mu_B^{\rm br} k }}{k^{1+\sigma}} \left(
    1  +   \frac{\tilde B}{k^{\theta_c}}  
   + \ldots \right)\,.
\end{align}

Now, we are ready to generalize this result to the case when both YLE and RW singularities are present, we obtain   
\begin{align}
    b_k = 
      \tilde A_{\rm YLE} \frac{e^{-\hat \mu_B{\rm YLE} k }}{k^{1+\sigma}} \left(
    1  +   \frac{\tilde B_{\rm YLE}}{k^{\theta_c}}  
    + \ldots \right)
    + \tilde A_{\rm RW} \frac{e^{-\hat \mu_B^{\rm RW} k }}{k^{1+\sigma}} \left(
    1  +   \frac{\tilde B_{\rm RW}}{k^{\theta_c}}  
   + \ldots \right) + {\rm c.c.}\,.
\end{align}
The coefficients $\tilde A_{\rm YLE, RW}$ and $\tilde B_{\rm YLE, RW}$ are generally complex numbers. 
Taking into account that ${\rm Im}\, \hat \mu_B^{\rm RW} =  \pi$, we have 
\begin{align}
\nonumber
    b_k &= 
      |\tilde A_{\rm YLE}| \frac{e^{-\hat \mu_r^{\rm YLE} k }}{k^{1+\sigma}} \left(
    \cos(\hat \mu_i^{\rm YLE} k + \phi^{\rm YLE}_a)  +   \frac{|\tilde B_{\rm YLE}|}{k^{\theta_c}}
    \cos(\hat \mu_i^{\rm YLE} k + \phi^{\rm YLE}_b)
    + \ldots \right) \\ &
    + |\hat A_{\rm RW}|  (-1)^k \frac{e^{-\hat \mu_r^{\rm RW} k }}{k^{1+\sigma}} \left(
    1  +   \frac{|\hat B_{\rm RW}|}{k^{\theta_c}}  
   + \ldots \right)\, ,
   \label{Eq:bks}
\end{align}
where $\phi_a$ and $\phi_b$ are phases due to non-trivial phases of $\tilde A^{\rm YLE}$ and $\tilde B^{\rm YLE}$ and trivial real factors were absorbed into $|\hat A_{\rm RW}|$ and $|\hat B_{\rm RW}|$. 

A few comments about the obtained results:
\begin{itemize}
    \item The coefficients, $b_k$, are exponentially sensitive to the imaginary values of the positions of the \YLE singularities and sensitive to the edge critical exponent $\sigma$. 
    \item The confluent critical exponent $\theta_c = \nu_c \omega = \frac{\sigma+1}{3} \omega$  is about $0.6$ ($\omega$ can be found in Ref.~\cite{Borinsky:2021jdb}) and thus leads to an appreciable suppression of the corresponding terms. It is safe to drop these corrections for a large enough order of $k$. 
    \item Consider a free fermion of mass $m$ and spin $1/2$. To the arbitrary order in the degeneracy (see e.g. Ref.~\cite{Kapusta:2023eix}) we have 
    \begin{equation}
        \hat n_b = \frac{2}{\pi^2} \hat m^2 \sum_k \frac{(-1)^{k+1}}{k} K_2(k \hat m) \sin(k \hat \mu)\,.
    \end{equation}
    The Fourier coefficients of this expansion is trivial to obtain. 
    Using the asymptotic expansion of the modified Bessel function, $K_2(x) \approx \left(\frac{\pi}{2x}\right)^{1/2} e^{-x}$, one finds 
    \begin{align}
        b_k \propto \frac{1}{k^{3/2}} e^{-k\hat m}\,.
    \end{align}
    That coincides with the leading order of the last term of Eq.~\eqref{Eq:bks} with the expected $\sigma=1/2$ and $\hat \mu^{\rm RW}_r = \hat m  $. 
\end{itemize}

Again, as for the Taylor coefficients, in QCD, it is not known at what order $k$ the asymptotic behavior sets in. A model calculation can explore this.   
Here, I consider the quark-meson model computed using FRG for the local potential approximation. This approximation leads to $\sigma_{\rm LPA} = 1/5$ (about a factor of two larger than expected, but a significant improvement from the mean-field value of $\sigma_{\rm MF} = 1/2$  ). 
The model is described in detail in Ref.~\cite{Skokov:2010wb}. For more information on the calculations, see Ref.~\cite{Bryant:2024twd}. 

\begin{figure}
    \centering
    \includegraphics[width=0.49\linewidth]{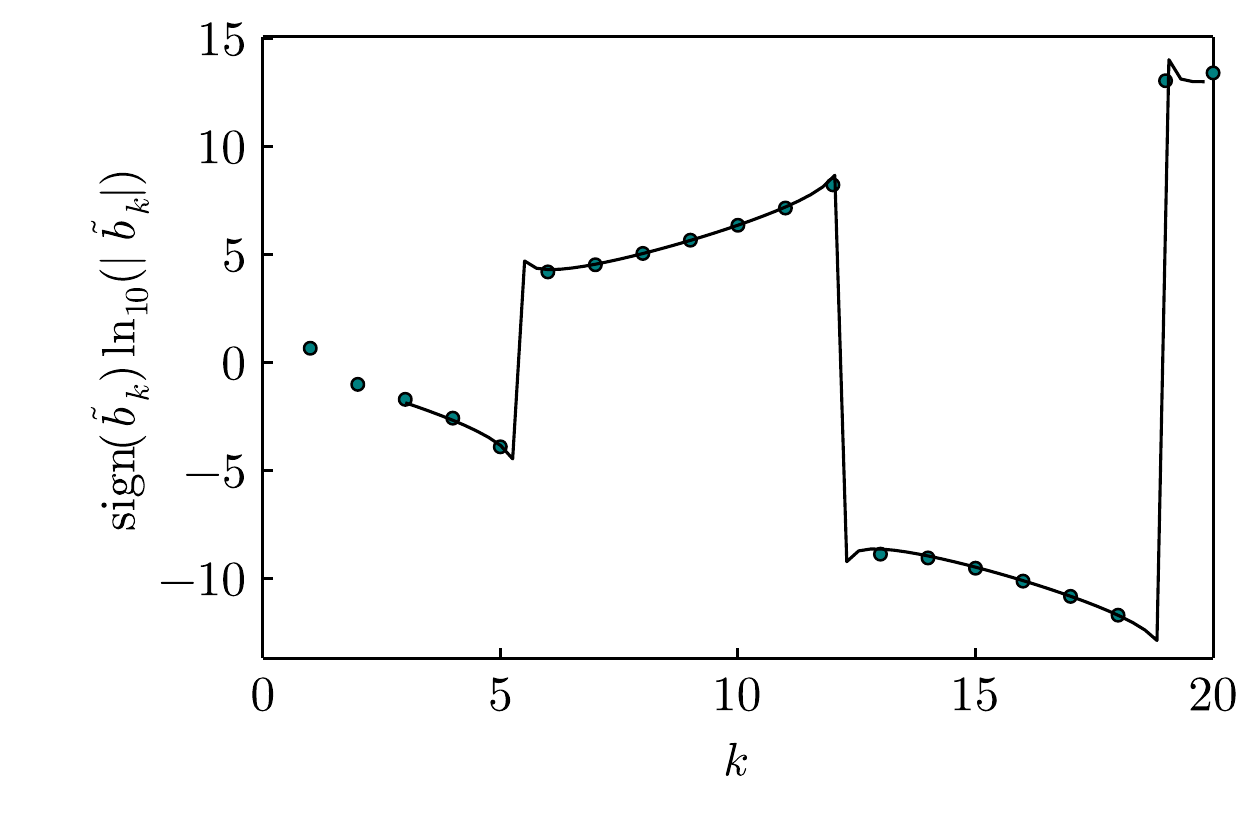}
    \includegraphics[width=0.49\linewidth]{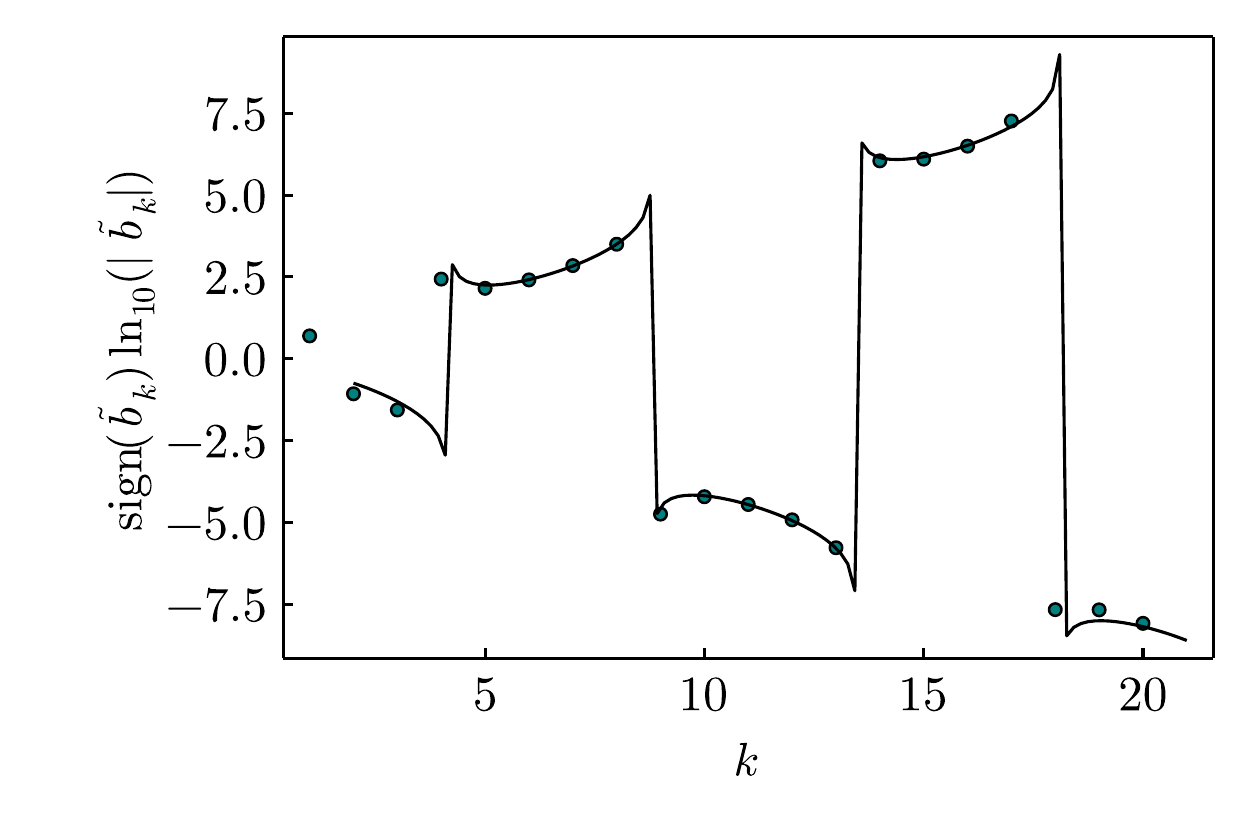}
    \caption{LPA FRG  Fourier coefficients $\tilde b_k = k^{1+\sigma} b_k$ ($\sigma_{\rm LPA} = 1/5$)  and the corresponding fits for $T = 150$ MeV (left) and $T = 180$ MeV (right).
    The figure is from Ref.~\cite{Bryant:2024twd}.}
    \label{fig:frg}
\end{figure}

Figure~\ref{fig:frg} 
shows the dependence of the Fourier coefficients on the order $k$ for two temperatures. The asymptotic  form  
Eq.~\eqref{Eq:bks} fits this dependence and extracts the location of the \YLE.  As one can see from the figure, the asymptotic behavior 
sets in for rather modest values of $k$. The fit yields $\hat \mu^{\rm fit}_{\rm YLE} = 1.483(7) + i\, 0.446(6)$ 
($\hat \mu^{\rm fit}_{\rm YLE} = 0.949(8) + i\, 0.675(11)$)
for $T = 150 (180)$  MeV.  The actual location of the singularity can be found using FRG, and it  is $\hat \mu_{\rm YLE} = 1.553 + i\, 0.4794$ ($\hat \mu_{\rm YLE} =  
0.9445 + i\,  0.6618$)  for $T = 150 (180)$  MeV. 
Although it was not possible to reliably extract the Fourier coefficient for $k>22$, 
the fit accuracy is sufficiently high as it reproduces the location within 5\% precision.   

The model results do not capture the first principle lattice QCD calculations, where the asymptotics might set in at larger values of $k$. Most importantly, model results provide high-precision data for $n_B(\mu_B)$, 
while lattice QCD results have unavoidable statistical uncertainty, which may significantly affect the extraction of $b_k$.

\section{Summary}
In these lecture notes, I provided an introduction to the analytic structure of the equation of state near a second-order phase transition with a specific focus on the \YLE singularity.
I reviewed the results of recent papers on the universal location of the  \YLE for different universality classes. In relation to QCD, I tried to show why understanding the analytic 
structure of QCD may be a key to locating the critical point and establishing the equation of state for non-zero values of the baryon chemical potential. I reviewed very recent results on the location of the critical point obtained through the analysis of the location of the \YLE extracted by the analytic continuation of the imaginary baryon chemical potential data and Taylor series coefficients computed on the lattice. Finally, 
I also discussed how the location of the \YLE and its properties manifest themselves in the asymptotics of baryon number Fourier coefficients and Taylor series coefficients.

\section*{Acknowledgements}
I am grateful to S. Tiwari for carefully reading these notes.   
I thank the organizers of the 2024 XQCD Ph.D. school and the conference  for their support 
and specifically  Yi Yin for hospitality.

\paragraph{Funding information}
This work 
is supported by the U.S. Department of Energy, Office 
of Nuclear Physics through contract DE-SC0020081.

\bibliography{bib}

\end{document}